\documentclass[amsmath,amssymb,twocolumn,superscriptaddress,pr,longbibliography]{revtex4-1}
\usepackage{physics}
\usepackage[normalem]{ulem}
\usepackage{amsfonts}
\usepackage{bm}
\usepackage{float}
\usepackage{graphicx}
\usepackage{xcolor}
\usepackage{verbatim}
\usepackage{mathtools}

\DeclarePairedDelimiter\floor{\lfloor}{\rfloor}
\usepackage{MnSymbol}
\usepackage{dsfont}
\usepackage[colorlinks=true, breaklinks=true, linkcolor=red, citecolor=blue, urlcolor=blue]{hyperref}

\begin{document}
\title{Steady-state dynamical mean field theory based on \\ influence functional matrix product states}
\author{Mithilesh Nayak}
\affiliation{Department of Physics, University of Fribourg, 1700 Fribourg, Switzerland}
\author{Julian Thoenniss}
\affiliation{Department of Theoretical Physics, University of Geneva, Quai Ernest-Ansermet 30, 1205 Geneva, Switzerland}
\author{Michael Sonner}
\affiliation{Max Planck Institute for the Physics of Complex Systems, 01187 Dresden, Germany}
\author{Dmitry A. Abanin}
\affiliation{Department of Physics, Princeton University, Princeton, New Jersey 08544, USA}
\affiliation{\'Ecole Polytechnique F\'ed\'erale de Lausanne (EPFL), 1015 Lausanne, Switzerland}
\author{Philipp Werner}
\affiliation{Department of Physics, University of Fribourg, 1700 Fribourg, Switzerland}

\begin{abstract}
We implement the recently developed influence functional matrix product states approach as impurity solver in equilibrium and nonequilibrium dynamical mean field theory (DMFT) calculations of the single-band Hubbard model. The method yields numerically exact descriptions of metallic states without sharp spectral features, at a moderate numerical cost. Systems with narrow quasi-particle or spin-polaron peaks, as well as low-temperature Mott insulators provide more challenges, since these simulations require long time contours or high bond dimensions. A promising field of application is the DMFT simulation of nonequilibrium steady states, which we demonstrate with results for photo-doped Mott systems with long-lived doublon and holon populations.  
\end{abstract}
\maketitle

\section{Introduction}

Dynamical mean field theory (DMFT) maps a lattice model onto an effective quantum impurity model with a self-consistently determined bath \cite{Georges1996}. This mapping, which neglects spatial correlations but retains time-dependent local fluctuations, is justified in the limit of high connectivity \cite{Metzner1989,MuellerHartmann1989}. For finite-dimensional lattices, DMFT is an approximation which  significantly reduces the numerical complexity and allows one to study bulk systems in parameter regimes where Monte Carlo simulations \cite{Blankenbecler1981} and other numerically exact approaches  are prohibitively expensive. The effective impurity models are, however, still nontrivial many-body systems whose solution requires dedicated numerical schemes. 

For equilibrium impurity models, there exists a range of powerful impurity solvers which provide access to low temperatures, even in the strongly correlated regime and in the case of multi-orbital and cluster impurities. In particular, the development of continuous-time quantum Monte Carlo (CTQMC) \cite{Rubtsov2005,Werner2006,Gull2011}, numerical renormalization group (NRG) \cite{Bulla2008,Weichselbaum2017} and tensor-network impurity solvers such as the dynamical density renormalization group (D-DMRG) \cite{Kuhner1999, Michal2008}, time-evolving block decimation (TEBD) \cite{Ganahl2015}, and real-frequency methods \cite{Ganahl2014, Bauernfeind2017} have greatly enhanced the scope of DMFT studies and enabled investigations of low-temperature electronic ordering phenomena \cite{Haule2007,Hoshino2016} and Hund metal crossovers \cite{Werner2008,Haule2009,Georges2013}. 

An active research frontier is the extension of DMFT and related methods such as GW+DMFT \cite{Biermann2003,Nielsson2017} to the nonequilibrium domain. On a formal level, this extension is rather straightforward and essentially involves the formulation of the self-consistency equations on a Keldysh or Kadanoff-Baym contour instead of the Matsubara axis \cite{Freericks2006,Aoki2014,Golez2019}. The limited availability of accurate and efficient nonequilibrium impurity solvers however represents a major hurdle. Weak-coupling perturbative solvers \cite{Tsuji2013}, which expand the self-energy in terms of the bare bath Green's functions, provide adequate solutions for the short-time dynamics in weakly correlated systems, but they do not conserve the total energy in isolated systems and become unstable in the intermediate or strong correlation regime,  where non-perturbative effects dominate. Boldified versions of such solvers, which use the interacting Green's functions instead of the bath Green's functions in the weak-coupling diagrams, can be derived from a Luttinger-Ward functional and hence are conserving. However, they have been found to strongly overestimate damping effects \cite{Eckstein2009}. The implementation of CTQMC on the Kadanoff-Baym contour yields the exact short-time DMFT dynamics, but the errors grow exponentially with the length of the real-time branches, due to a dynamical sign problem \cite{Muehlbacher2008,Werner2009}. The self-consistent strong-coupling expansion \cite{Keiter1971,Grewe1981,Pruschke1989,Eckstein2010nca}, which starts from the atomic limit and treats the hybridization term as a perturbation, has been intensively used for studies of nonequilibrium phenomena in Mott systems \cite{Eckstein2010breakdown,Murakami2018steady,Murakami2018hhg}. The cost of these calculations, however, increases steeply with increasing diagram order and practical applications have been mostly limited to the first-order (noncrossing) \cite{Keiter1971} and second-order (one-crossing) \cite{Pruschke1989} approximations. Especially for metallic systems, or multiorbital impurities, these low-order approaches are not accurate enough. 

In recent years, some promising strategies have emerged which provide a realistic path towards numerically exact nonequilibrium DMFT simulations. One of them is the inchworm Monte Carlo approach \cite{Cohen2015}, which stochastically samples the self-energy diagrams of the self-consistent strong coupling expansion \cite{Keiter1971,Bickers1987}. Since this expansion rapidly converges in the strong correlation regime, essentially exact results can be obtained in some relevant situations at relatively low expansion orders, which are manageable even in the presence of a sign problem \cite{Kuenzel2024}. The development of the tensor cross interpolation (TCI) technique \cite{Oseledets2010,Fernandez2022} enables an efficient and noise-free calculation of these low-order diagrams, and is likely to replace the Monte Carlo approach. Both inchworm and TCI-based strong coupling solvers have recently been implemented for nonequilibrium steady-state setups \cite{Kuenzel2024,Kim2024}.   
A downside of the self-consistent strong-coupling expansion, which is based on the pseudoparticle formalism \cite{Bickers1987,Eckstein2010nca}, is that certain pseudoparticle Green's functions decay very slowly. This makes it necessary to introduce damping factors and treat long time intervals in steady-state simulations \cite{Li2021}. 

An alternative method, which works with physical Green's functions, is the influence functional (IF) approach \cite{Feynman1963}.  For noninteracting baths, the IF is a $(0+1)$-dimensional Gaussian functional that encodes bath-induced memory effects on the impurity dynamics. Recent studies showed that the IF can be efficiently approximated by a matrix product state (MPS) with moderate bond dimensions \cite{Thoenniss2023a,Thoenniss2023b,ng2023real,Chen2024a,Chen2024b}. In the case of nonequilibrium quantum dot problems, this approach reproduces previously obtained inchworm Monte Carlo results with moderate computational effort \cite{Thoenniss2023b}. 

Here, we implement nonequilibrium DMFT simulations using an IF matrix product state (IF-MPS)  impurity solver. We show that this approach produces accurate real-frequency spectral functions for equilibrium systems in moderately correlated metallic states and enables an efficient simulation of quasi-steady states in photo-doped Mott insulators. The IF-MPS approach thus promises to become one of the practically useful methods for numerically exact nonequilibrium DMFT calculations.      

The rest of this paper is organized as follows. In Sec.~\ref{sec_method}, we detail the implementation of the IF method on the Keldysh contour. In Sec.~\ref{sec_impurity}, we present some benchmarks for an impurity setup, while Sec.~\ref{sec_results} demonstrates the convergence of the results as a function of truncation and discretization parameters, and shows DMFT results for equilibrium and nonequilibrium setups. Section~\ref{sec_conclusions} is a brief conclusion and outlook. 

\section{Model and method}
\label{sec_method}

\subsection{Model}

We introduce the IF-MPS impurity solver for the case of the single impurity Anderson model (SIAM) \cite{Anderson1961}. The SIAM is described by the Hamiltonian
\begin{eqnarray}
H &=&H_{\mathrm{imp}} + \sum_{\sigma, p}  \omega_{p, \sigma}c^{\dagger}_{p,\sigma}c_{p,\sigma} + \sum_{\sigma, p} (V_{p,\sigma}d^\dagger_{\sigma}c_{p,\sigma} + \mathrm{H.c.}),\label{eqn_SIAM}\hspace{5mm}\\
H_{\mathrm{imp}} &=& (\epsilon_{d} - U/2)\sum_{\sigma} n_\sigma +U n_{\uparrow}n_{\downarrow},\label{eq_Himp}
\end{eqnarray}
where $c^{\dagger}_{p,\sigma}$ $(c_{p,\sigma})$ denotes the fermionic creation
(annihilation) operator of the bath mode indexed by a quantum number $p$ and
spin $\sigma=\uparrow,\downarrow$, and $d^{\dagger}_{\sigma}$ $(d_{\sigma})$
are the fermionic creation (annihilation) operators of the impurity states with
spin $\sigma$. 
In Eq.~\eqref{eq_Himp}, $n_\sigma=d^\dagger_\sigma d_\sigma$ is the spin-density on the impurity site. 
The bath energy levels are $\omega_{p,\sigma}$ and $V_{p,\sigma}$ denotes the hopping amplitude between the impurity and the
$\sigma$-spin bath mode $p$. The impurity potential is $\epsilon_{d}$ and $U$ is the Hubbard
repulsion between electrons with opposite spins on the impurity site.
$\epsilon_d=0$ corresponds to a half-filled impurity for any $U$. 
 
\subsection{Influence functional formalism}
\label{IF_formalism}

\subsubsection{Correlation functions}

The task of an impurity solver is to compute local observables and temporal correlation functions on the impurity site. We define the real-time two-point correlation function of the operators $O_1$ and $O_2$ at times $t_2\ge t_1\ge 0$ as
\begin{eqnarray}
 \langle O_2(t_2) O_{1}(t_1)\rangle_\text{imp}=  \mathrm{Tr}\lbrack e^{iHt_2} O_{2} e^{-iH(t_2-t_1)} O_{1} e^{-iHt_1} \rho \rbrack,\hspace{2mm}\hspace{3mm}\label{eq_correlator}
\end{eqnarray} 
where the density matrix $\rho$ denotes the initial state at $t=0$. We are mostly interested in steady states, where the density matrix is conserved, $\rho_\mathrm{ss}=e^{-iHt}\rho_\mathrm{ss} e^{iHt}$, and the two-point correlation functions are time-translation invariant, 
\begin{eqnarray}
 \langle O_2(t_2) O_{1}(t_1)\rangle_\text{imp} = \langle O_2(t_2-t_1) O_{1}(0)\rangle_\text{imp}.
\end{eqnarray}
However, in the IF-MPS approach, the time evolution starts with a product initial density matrix for the impurity and the bath, 
\begin{eqnarray}
 \rho(0) = \rho_\mathrm{imp} \otimes \rho_\mathrm{bath}.\label{Eq_factorization}
\end{eqnarray}
In many physical systems, the local dynamics relaxes quickly with some
equilibration timescale $t_\mathrm{relax}$ if the bath is large and in a
thermal state. We can therefore compute the stationary two-point functions by
choosing $t_1\gg t_\mathrm{relax}$, and carefully monitoring convergence with
$t_1$.

We define a maximal evolution time $t_\mathrm{max}$ such that
$0 \le t_1 \le t_2 \le t_\mathrm{max}$. To obtain the Keldysh path-integral representation
of Eq.~(\ref{eq_correlator}), we discretize the time interval
$[0,t_\text{max}]$ into $M$ slices of length $\delta t = t_\text{max}/M$ with $t_i=m_i\delta t$,
and define $U_{\delta t} = e^{-iH\delta t}$, so that 
\begin{eqnarray}\label{eq:trace_expression}
\langle O_2(t_2) O_{1}(t_1)\rangle_\text{imp} &=&  \mathrm{Tr}\Big[U_{\delta t}^{-M} U_{\delta t}^{M - m_2}{O}_{2}U_{\delta t}^{m_2-m_1}{O}_{1} U_{\delta t}^{m_1}\rho\Big],\nonumber\\
\end{eqnarray} 
and then introduce a resolution of the Grassmann identity \cite{Negele1988},
\begin{equation}
\label{eq:simple_GM_identity}
\mathds{1} = \int d\bar{\eta}d\eta\, e^{-\bar{\eta}\eta}|\eta\rangle\langle\bar{\eta}|,
\end{equation} 
between successive time-evolution operators for every degree of freedom, with
the Grassmann coherent states $|\eta\rangle$ and their conjugate $\langle
\bar{\eta}|$ defined as
\begin{align}\label{eq:coherent_state1}
\ket{\eta} = e^{-\eta d^\dagger}\ket{\emptyset},\\
\label{eq:coherent_state2}
\bra{\bar{\eta}} = \bra{\emptyset} e^{\bar{\eta} d}.
\end{align}
For non-interacting environments, the bath degrees of freedom can be integrated out exactly. In the continuous-time limit $\delta t\to 0$, this yields the standard $(0+1)$-dimensional field-theory description of the impurity dynamics~\cite{Negele1988}:
\begin{eqnarray}\label{eq:standard_path_integral}
&&\langle O_2(t_2)O_{1}(t_1)\rangle_\text{imp} = \int \left(\prod_{\sigma, z}d\bar{\eta}_{\sigma, z}d\eta_{\sigma, z}\right)\nonumber\\
&&\hspace{2mm}\times ~\mathcal{O}_2[\bar{\eta}_{\uparrow, t_2^+}, \bar{\eta}_{\downarrow, t_2^+}, \eta_{\uparrow, t_2^+},\eta_{\downarrow, t_2^+} ]\mathcal{O}_1 [\bar{\eta}_{\uparrow, t_1^+}, \bar{\eta}_{\downarrow, t_1^+}, \eta_{\uparrow, t_1^+},\eta_{\downarrow, t_1^+} ]\nonumber\\
&&\hspace{2mm}\times ~\mathrm{exp}\left\lbrace \int_{\mathcal{C}}dz\left\lbrack \sum_{\sigma}\bar{\eta}_{\sigma, z}\partial_z\eta_{\sigma, z} - i \mathcal{H}_{\mathrm{imp}}[\bar{\eta}_{\uparrow, z}, \bar{\eta}_{\downarrow, z}, \eta_{\uparrow, z},\eta_{\downarrow, z}]\right\rbrack\right\rbrace\nonumber\\
&&\hspace{2mm}\times\prod_{\sigma}\mathrm{exp}\left(\int_{\mathcal{C}}dz\int_{\mathcal{C}}dz'~\bar{\eta}_{\sigma, z}\Delta_\sigma(z,z')\eta_{\sigma, z'}\right) \rho_\text{imp}[\bar{\eta}_{\sigma,0},\eta_{\sigma,0}].\nonumber\\
\end{eqnarray}
Here, $\mathcal{C}=0\rightarrow t_\text{max} \rightarrow 0$ 
denotes a real-time contour with forward and backward branches, $z\in\mathcal{C}$ is a contour time variable, and $\bar \eta_{\sigma,z}$ and $\eta_{\sigma,z}$ are Grassmann variables that parametrize the impurity trajectory. In this representation, the bath's effect
enters through the hybridization function, whose spectral representation is given by
\begin{eqnarray}
\Delta_{\sigma}(\omega) = \sum_p\frac{|V_{p,\sigma}|^2}{\omega-\omega_{p,\sigma}},
\end{eqnarray}
with $V_{p,\sigma}$ and $\omega_{p,\sigma}$ as described in Eq.~\eqref{eqn_SIAM}.  

Evaluating Eq.~(\ref{eq:standard_path_integral}) for generic AIMs is challenging due to (i) interacting dynamics that resists exact solutions and perturbative treatments, and (ii) time non-locality, which precludes the use of master-equation approaches and standard time-evolution techniques. The IF-MPS approach employed in this work overcomes these challenges by restoring time-locality in Eq.~(\ref{eq:standard_path_integral}). This is achieved by representing the time-nonlocal correlations encoded by the hybridization function as a matrix product state, where each tensor functions as an effective super-evolution operator acting on a combined impurity-bath system. In the following, we briefly review the method before detailing its integration with DMFT.

\subsubsection{Influence Functional}

The Gaussian functional in the last line of Eq.~(\ref{eq:standard_path_integral}) is the continuous-time version of the Feynman-Vernon influence functional (IF). 
In order to obtain an IF representation with a finite number of degrees of freedom, we work with time steps $\delta t$. Furthermore, we separate the bath evolution from the impurity evolution by Trotterizing the evolution operator: 
\begin{equation}\label{eq:Trotter}
U_{\delta t} =  U_{\text{imp},\delta t} \cdot U_{\text{hyb},\delta t} +\mathcal{O}(\delta
t^2),\end{equation} with $U_{\text{imp},\delta t} = e^{-iH_\mathrm{imp},\delta t}$ and $U_{\text{hyb},\delta t} = e^{-i(H-H_\mathrm{imp})\delta t}.$
To derive the discrete-time path integral, one thus needs to insert $4M$ Grassmann coherent states per spin species into Eq.~(\ref{eq:trace_expression})---two per Trotterized time step on the forward and backward branch, respectively.  As each identity resolution introduces two Grassmann variables (see Eq.~(\ref{eq:simple_GM_identity})), the discrete-time version of Eq.~(\ref{eq:standard_path_integral}) is parametrized by $8M$ impurity Grassmann variables per spin species. Half of these ($4M$) parametrize the IF and the other half parametrize the local impurity gates. 
We collect the IF-variables in a vector (dropping the spin label $\sigma$  for clarity):  \begin{align}\label{eq:GM_vec}
\boldsymbol{\eta} &\equiv (\boldsymbol{\eta}_0,\boldsymbol{\eta}_1,\dots, \boldsymbol{\eta}_{M-1}),\\
\boldsymbol{\eta}_m &\equiv ( \eta_{2m^+},\eta_{2m^-},\eta_{2m+1^+},\eta_{2m+1^-} ), \end{align}
with $m\in \{0,1,\dots, M-1\}.$ The superscripts ``$+$'' and ``$-$'' denote the forward and backward Keldysh branch, respectively.
 At leading order in $\delta t$, the time-discrete IF in Eq.~(\ref{eq:standard_path_integral}) then takes the form 
\begin{eqnarray}
\mathcal{I}_\sigma[\boldsymbol{\eta}_{m}] = \mathrm{exp}\left(\frac{1}{2}~\sum_{m,n}\boldsymbol{\mathbf{\eta}}^{\mathrm{T}}_{m}~ \mathbf{B}^\sigma_{m,n}\boldsymbol{\mathbf{\eta}}_{n}\right).
\label{eq_defn_IF}
\end{eqnarray} 
Here, $\mathbf{B}_{m,n}^\sigma$ is a $4\times 4$ matrix with elements \cite{Thoenniss2023b},
$$
\mathbf{B}^\sigma_{m,n} = \delta_{m,n}\begin{bmatrix}0&0&-1&0\\ 0&0&0&1\\1&0&0&0\\0&-1&0&0\end{bmatrix} - \mathbf{\Delta}^\sigma_{m,n}{(\delta t)}^2.
$$
It encompasses the overlaps of the Grassmann coherent states (first term) and
the contribution from the discretized hybridization function
$\mathbf{\Delta}^\sigma_{m,n}$ (second term), 
which
is given by
\begin{eqnarray}
\mathbf{\Delta}^\sigma_{m,n} = \int d\omega A_\text{bath}(\omega)\mathbf{G}^\sigma_{m,n}(\omega; \beta, \mu) +\mathcal{O}(\delta t).
\label{Delta_defn}
\end{eqnarray}
In this expression, $A_\text{bath}(\omega)$ is the bath density of states
(DOS). The Green's function matrix $\mathbf{G}_{m,n}$ is given by (see
Appendix~\ref{app_A}) \footnote{In the derivation of the influence functional, we define the Green's functions  
without a $-i$
factor, to be consistent with the previous literature \cite{Thoenniss2023a,Thoenniss2023b}.}

\begin{eqnarray}
\mathbf{G}^\sigma_{m>n} = \begin{pmatrix}
0 & g^{>,*}_{m,n} & g^{<,*}_{m,n} & 0 \\
-g^{>}_{m,n} &0 &0 &-g^{<}_{m,n}\\
-g^{>}_{m,n} &0 &0 &-g^{<}_{m,n}\\
0 & g^{>,*}_{m,n} & g^{<,*}_{m,n} &0  
\end{pmatrix},
\label{Green's_function_matrix_defn1}
\end{eqnarray}
and 
\begin{eqnarray}
\mathbf{G}^\sigma_{m=n} = ~\frac{1}{2}~\begin{pmatrix}
0 & g^{>,*}_{m,m} & g^{>}_{m,m} & 0 \\
-g^{>}_{m,m} &0 &0 &-g^{>,*}_{m,m}\\
-g^{>}_{m,m} &0 &0 &-g^{<}_{m,m}\\
0 & g^{>,*}_{m,m} & g^{<,*}_{m,m} &0  
\end{pmatrix},
\label{Green's_function_matrix_defn2}
\end{eqnarray}
where
\begin{eqnarray}
g^<_{m,n}(\omega; \beta, \mu) &=& -f_{F}(\omega; \beta, \mu)e^{-i\omega(m-n)\delta t},\label{nonint_les}\\
g^>_{m,n}(\omega; \beta, \mu) &=& \left(1-f_{F}(\omega; \beta, \mu)\right)e^{-i\omega(m-n)\delta t}\label{nonint_gtr}
\end{eqnarray}
denote the lesser and greater Green's function for a noninteracting bath electron with energy level $\omega$, chemical potential $\mu$ and inverse temperature $\beta$; $f_F(\omega; \beta, \mu)=1/\left(1+e^{\beta(\omega-\mu)}\right)$ is the Fermi-Dirac distribution
function for this temperature. 

\begin{figure*}
\centering
\includegraphics[width = 18cm, height=18cm, keepaspectratio]{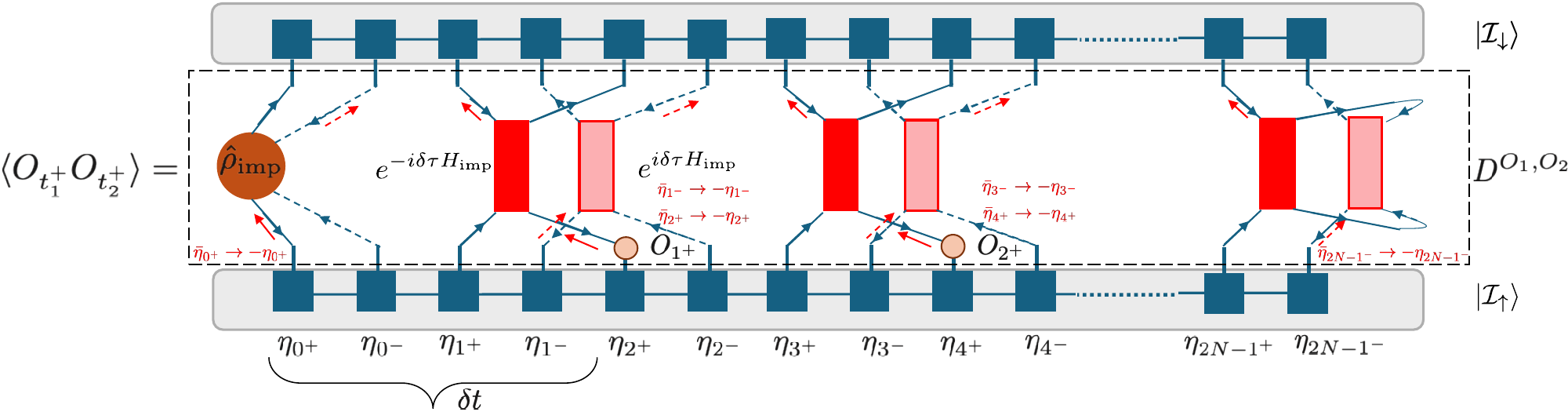}
\caption{Two-point correlators can be evaluated by a tensor network contraction of two IF-MPS corresponding to $|I_\uparrow\rangle$ and $|I_\downarrow\rangle$ (shown in grey boxes), and the matrix product operator $D^{O_1 O_2}$ (dashed box). The blue arrows show the flow of time along the forward and backward branches of the Keldysh contour, while the red arrows represent the variable transformations \eqref{eq:var_sub1} and \eqref{eq:var_sub2}. The corresponding signs are absorbed into the matrix product operator.
In the curly braces we show a full time-evolution step $\delta t$ on the forward and backward branches.}
\label{IFMPS_Tensor_network}
\end{figure*}

\subsubsection{Expectation Value as State Overlap}

Having defined the discrete-time IF in Eq.~(\ref{eq_defn_IF}), we return to evaluating the expectation value in Eq.~(\ref{eq:standard_path_integral}).  Upon making appropriate variable substitutions that introduce signs on the impurity gates while leaving the IF from Eq.~(\ref{eq_defn_IF}) unchanged (see App.~\ref{app:overlap_derivation} for details), Eq.~(\ref{eq:standard_path_integral}) can be expressed as:
\begin{align}
&\langle O_2(t_2)O_{1}(t_1)\rangle_\text{imp}  \;  = \int \bigg(\prod_{\sigma} d\bar{\bm{\eta}}_{\sigma} d\bm{\eta}_{\sigma}\bigg) \nonumber\\
& \times \mathcal{I}_\downarrow[\bm{\eta}_\downarrow]  e^{-\bar{\bm{\eta}}_\downarrow \bm{\eta}_\downarrow} \mathcal{D}^{O_1O_2}[\bar{\bm{\eta}}_\downarrow, \bm{\eta}_\uparrow] 
e^{-\bar{\bm{\eta}}_\uparrow \bm{\eta}_\uparrow} \mathcal{I}_\uparrow[\bar{\bm{\eta}}_{\uparrow}] + \mathcal{O}(\delta t),
\label{eq:expec_value_overlap}
\end{align}
where $\mathcal{D}^{O_1O_2}[\bar{\bm{\eta}}_\downarrow, \bm{\eta}_\uparrow]$ denotes the Grassmann kernel that encodes the local impurity evolution operator $U_{\text{imp},\delta t}$, the operators $O_1(t_1 = m_1\delta t)$ and
$O_2(t_2 = m_2\delta t)$, and the initial impurity density matrix $\rho_\text{imp}$. \\

 By replacing the Grassmann variables by creation operators
 $\bm{d}_m^\dagger\equiv
 (d^\dagger_{2m^+},d^\dagger_{2m^-},d^\dagger_{2m+1^+},d^\dagger_{2m+1^-})$ in
 Eq.~(\ref{eq_defn_IF}) using the definitions of the coherent states in
 Eqs.~(\ref{eq:coherent_state1}--\ref{eq:coherent_state2}), one obtains the
 state representation of the IF to leading order in $\delta t$:
\begin{equation}
    \ket{I_\sigma} = \exp\Big( \frac{1}{2}\sum_{m,n}{\mathbf{d}_m^\dagger}^T\mathbf{B}_{m,n}^\sigma\, \mathbf{d}^\dagger_n\Big)\ket{\emptyset}.
    \label{eq:Istate}
\end{equation}
Such a state formally has a Bardeen-Cooper-Schrieffer form, regardless of the
fermion-number conservation of the original problem. Physical symmetries are
reflected in the structure of the matrix $\mathbf{B}.$

Similarly, the Grassmann kernel $\mathcal{D}^{O_1O_2}$ can be represented as a fermionic many-body operator (see App.~\ref{app:overlap_derivation}), which we denote by $D^{O_1O_2}$. With this, Eq.~(\ref{eq:expec_value_overlap}) can be cast into the form 
\begin{equation}
\label{eq:state_overlap}
\langle O_2(t_2)O_{1}(t_1)\rangle_\text{imp}   = \langle I_\downarrow | D^{O_1O_2}|I_\uparrow \rangle.
\end{equation}

Our strategy is to evaluate this overlap as a tensor contraction, where the 
IF-states  $\ket{I_\sigma}$ are represented as matrix product states and the operator $D^{O_1O_2}$ is represented as a matrix product operator (MPO), see Fig.~\ref{IFMPS_Tensor_network}.   The tensors of this MPO encode 
(i) the unitary impurity evolution operators \( U_{\text{imp},\delta t} \),  
(ii) the operators \( O_1 \) and \( O_2 \), and  
(iii) the initial impurity density matrix \( \rho_{\text{imp}} \).  
For the single-orbital model considered here, the MPO has a trivial bond dimension \( \chi = 1 \) due to the time-locality of the local impurity evolution \footnote{As is common in the tensor network literature, the object $D^{O_1O_2}$ is shown in Fig.~\ref{IFMPS_Tensor_network} as a tensor product of impurity evolution gates.}. The derivation of its effective tensors---which differ from standard unitary gates as they appear in conventional tensor time evolution methods---is detailed in Sec.~\ref{sec:operator_rep_imp} and App.~\ref{app:impurity_gate}.

\subsubsection{MPS Representation of the IF-state 
}\label{sec:MPS_IF}

Recently, multiple paths of constructing MPS representations of many-body states in the form of Eq.~(\ref{eq:Istate}) were explored~\cite{Thoenniss2023a,Thoenniss2023b,ng2023real,Chen2024a,Chen2024b}. In this work, we follow Refs.~\cite{Thoenniss2023a,Thoenniss2023b}, which generalize an algorithm that was originally introduced by Fishman and White (FW) \cite{Fishman2015}.  While Ref.~\cite{Thoenniss2023a} contains a detailed derivation, we summarize its key steps here for completeness. The central idea of this approach is to find a sequence of Gaussian two-body rotations which map the IF state onto the vacuum state at the \emph{single-body level}. By expressing this circuit of rotation matrices as unitary \emph{many-body gates} and applying them in reverse order to a MPS representation of the many-body vacuum state, one obtains the many-body state $\ket{I_\sigma}$. At the single-body level, we start by expressing the Gaussian IF-state \( \ket{I_\sigma} \) which is fully characterized by its two-point functions, in terms of its correlation matrix
\begin{eqnarray}
\label{eq:corr_matrix}
\Lambda^\sigma_{\tau,\tau^\prime} = 
\begin{bmatrix}
\langle d_{\tau}d^\dagger_{\tau^\prime}\rangle_{I_\sigma} &\langle d_{\tau} d_{\tau^\prime}\rangle_{I_\sigma}  \\
\langle d^\dagger_{\tau}d^\dagger_{\tau^\prime}\rangle_{I_\sigma} &\langle d^\dagger_{\tau} d_{\tau^\prime}\rangle_{I_\sigma} 
\end{bmatrix}.
\end{eqnarray} 
Here, 
\begin{equation} 
\langle d_{\tau} d^\dagger_{\tau^\prime}\rangle_{I_\sigma}  \equiv \frac{\bra{I_\sigma} d_{\tau} d_{\tau^\prime}^\dagger\ket{I_\sigma}
}{\bra{I_\sigma}I_\sigma\rangle}, \label{eq:corr_Isigma}
\end{equation} 
and analogously for the other
components. We stress that the two-point correlations $\langle \dots \rangle_{I_\sigma} $ in Eq.~(\ref{eq:corr_matrix}) merely encode the correlations of the IF defined in Eq.~(\ref{eq:Istate}). In contrast to the physical expectation value $\langle \dots\rangle_\text{imp}$ used in  Eqs.~(\ref{eq_correlator}--\ref{eq:standard_path_integral}), they do not contain any physical information about the impurity dynamics and are only needed to construct the MPS-representation of $\ket{I_\sigma}$. \\
In Eqs.~\eqref{eq:corr_matrix} and \eqref{eq:corr_Isigma}, we introduced the time index $\tau\in \mathcal{C}$, 
which encompasses a time step index and a branch label $\alpha\in\{+,-\}$, i.e. $\tau = m^\alpha\in\{0^+,0^-,1^+,1^-,\dots,(2M-1)^+, (2M-1)^-\}$.  Furthermore, with each $\tau$ we associate an integer index according to \begin{equation}
|\tau| = \begin{cases} 2m &  \text{ if } \alpha = +, \\
2m+1 &\text{ if } \alpha = -.
\end{cases}
\end{equation}
The entries in Eq.~(\ref{eq:corr_matrix}) are then evaluated as (see Ref.~\cite{Caracciolo2013Algebraic}, Eq.~(A.90)):
\begin{align}
\langle d^\dagger_{\tau} d_{\tau^\prime}\rangle_{I_\sigma}  &= \text{pf}\Big[\big(\tilde{\mathbf{B}}_\sigma^{-1}\big)^{T} \Big|_{[i_1,i_2][i_1,i_2]}\Big],\\
\langle d_{\tau} d_{\tau^\prime}\rangle_{I_\sigma}  &= \text{pf}\Big[\big(\tilde{\mathbf{B}}_\sigma^{-1}\big)^{T} \Big|_{[i_2,i_2][i_2,i_2]}\Big],\\
\langle d^\dagger_{\tau} d^\dagger_{\tau^\prime}\rangle_{I_\sigma}  &= \text{pf}\Big[\big(\tilde{\mathbf{B}}_\sigma^{-1}\big)^{T}  \Big|_{[i_1,i_1][i_1,i_1]}\Big],\\
\langle d_{\tau} d^\dagger_{\tau^\prime}\rangle_{I_\sigma}   &=\delta_{\tau,\tau^\prime} -  \langle d_{\tau^\prime} d^\dagger_{\tau}\rangle_I ,
\end{align}
where the Pfaffian is computed for the $2\times2$ matrices determined by the
row and column indices indicated in the square brackets, with
\begin{align}
& i_1(\tau) = 4M + |\tau|, \quad i_2(\tau) = 8M + |\tau|.
\end{align}
In the above expectation values we introduced the $16M\times 16M$ matrix
\begin{equation}
\tilde{\mathbf{B}}_\sigma 
= \frac{1}{2} \begin{pmatrix}
(\mathbf{B}^\sigma)^\dagger & \mathds{1} & 0 & 0 \\
-\mathds{1}  & 0 & \mathds{1}  & 0 \\
0 & -\mathds{1}  & 0 & \mathds{1} \\
0  & 0& -\mathds{1}  & \mathbf{B}^\sigma
\end{pmatrix}.
\end{equation}
The problem of finding a sequence of two-site rotations mapping the
IF-state to the vacuum state is now equivalent to the problem of finding
rotations which diagonalize the correlation matrix. Since \(
\Lambda^\sigma_{\tau,\tau^\prime} \) represents a pure state, its diagonal form
reads
\begin{equation}
\Lambda^\sigma_{\tau,\tau^\prime} \xrightarrow{\text{diag.}} \begin{bmatrix}
  \langle \tilde{d}_{\tau}\tilde{d}^\dagger_{\tau^\prime}\rangle_{I_\sigma}  &\langle \tilde{d}_{\tau} \tilde{d}_{\tau^\prime}\rangle_{I_\sigma}  \\
  \langle \tilde{d}^\dagger_{\tau}\tilde{d}^\dagger_{\tau^\prime}\rangle_{I_\sigma} &\langle \tilde{d}^\dagger_{\tau} \tilde{d}_{\tau^\prime}\rangle_{I_\sigma}  
\end{bmatrix} =\begin{bmatrix}
\mathds{1} &0\\
0&0 
\end{bmatrix},
\end{equation} 
which corresponds to the vacuum state. 

The FW algorithm  performs the diagonalization step approximately: Starting from the
first fermionic mode associated with the operator pair
$d^\dagger_{\tau},d_{\tau}$, which corresponds to the first time step, the
fermionic modes are added until the corresponding submatrix of the correlation
matrix has an eigenvalue close to zero or one (up to a tolerance which we fixed
to $10^{-12}$) or a maximum number $n_\mathrm{sub}$ is reached. The
eigenvector of this eigenvalue corresponds to a single-body wavefunction with
definite occupation (associated with the operator pair $\tilde{d}^\dagger_{\tau},\tilde{d}_{\tau}$) \footnote{The entanglement entropy (E.E.) of the bipartition $\lbrack 0, n_{\mathrm{sub}}\rbrack $ and $\lbrack n_{\mathrm{sub}}+1, 4N \rbrack$ can be evaluated and one will have isolated a single fermion mode when the E.E is zero. The entanglement entropy of a single fermion reduced density matrix corresponding to the bipartition $\lbrack 0, n_{\mathrm{sub}}\rbrack$ can be evaluated from the two-point correlation matrix and is given by  $S(e) = - e\ln e - (1 - e)\ln (1-e)$, where $e$ is the eigenvalue of the sub-correlation matrix ($0\leq e\leq 1$) \cite{Fishman2015}.}.  
We then add a sequence of up to $n_\mathrm{sub}$ two-site rotations which rotate this mode of definite occupation to the edge. These steps are repeated starting from the second fermionic mode until we obtain an approximate diagonalization of the entire state. Finally, this sequence of rotations is applied in reverse on a MPS representation of the many-body vacuum state. In order to apply the many-body gates, we use a version of the time evolving block decimation~\cite{vidal2003efficient,vidal2004efficient}, with a given maximal bond dimension $\chi$ and a singular value decomposition (SVD) cutoff of $10^{-8}$.\par  

The maximum number of modes considered in each step, $n_{\mathrm{sub}}$, represents a memory time cutoff. The FW cutoff of $10^{-12}$ is an \emph{implicit} memory time cutoff which depends on the temporal correlation decay, whereas $n_\mathrm{sub}$ is a hard memory time cutoff $t_\mathrm{cutoff} = \frac{1}{4} n_\mathrm{sub}\delta t$ and must be carefully chosen with respect to the bond dimension $\chi$. If it is too large, more potentially irrelevant information about long-time correlations will be encoded within the finite bond dimension $\chi$, reducing the accuracy of the influence functional. Additionally, the deeper circuit leads to an increase of the numerical cost of evaluating the resulting quantum circuit, due to a larger number of SVD operations [complexity of the order of $\mathcal{O}(n_{\mathrm{sub}}M\chi^3)$]. If it is chosen to be too small, long-time correlations are truncated, which potentially reduces the accuracy as well. We empirically verify that the subsystem size of  
\begin{equation}
\label{eq:nsub_to_bond_dimension}
n_{\mathrm{sub}} = 4\floor*{\ln_2\chi}
\end{equation}
is a suitable choice (see Sec.~\ref{sec_impurity}), which is consistent with the single-fermion Gaussian formulation of the algorithm in Ref.~\cite{Fishman2015}.

We emphasize that the ordering of variables in the correlation matrix, Eq.~(\ref{eq:corr_matrix}), determines the order of the degrees of freedom in the IF-MPS. Choosing their order as in Eq.~(\ref{eq:GM_vec}) is crucial for the MPS to have a low bond dimension~\cite{banuls2009matrix,
hastings2015connecting, lerose2023overcoming}.  At the level of the product operator $D^{O_1O_2}$ in Eq.~(\ref{eq:state_overlap}), this variable ordering must be accounted for by adjusting the signs of each local impurity gate. This is explained in the following section.

\subsubsection{MPO Representation of the Impurity Operator $D^{O_1O_2}$}\label{sec:operator_rep_imp}

Expanding the IF and impurity operator appearing in Eq.~(\ref{eq:state_overlap}) in the full many-body basis $\{|\mu\rangle \}$, 
\begin{align}
    |I\rangle &= \sum_\mu I_\mu |\mu\rangle,\\
    D^{O_1O_2} &= \sum_{\mu,\nu} |\nu\rangle \alpha_{\nu,\mu} \langle \mu|,\label{eq:gate_coeffs}
\end{align} allows to associate each term with a specific impurity trajectory $|\mu\rangle.$ Here,
\begin{flalign}
    |\mu\rangle &\equiv d^\dagger_{i} d^\dagger_{j} \dots d^\dagger_{k}|\emptyset\rangle,\\
    \langle \mu| &\equiv \langle \emptyset| d_k \dots d_j d_i,
\end{flalign}
where $i<j<\dots <k$ with respect to the ordering in Eq.~(\ref{eq:GM_vec}), and analogously for $\ket{\nu}$ and $\bra{\nu}$. With this, Eq.~(\ref{eq:state_overlap}) can be expressed as
\begin{align}
 \bra{I_\downarrow }{D}^{O_1O_2} \ket{I_\uparrow} &= \sum_{\mu,\nu}\,I_{\downarrow,\nu} I_{\uparrow,\mu}\,  \langle \nu| \big(| \nu \rangle \alpha_{\nu,\mu} \, \langle \mu | \big) |\mu\rangle\nonumber\\
 &=\sum_{\mu,\nu}\,I_{\downarrow,\nu} I_{\uparrow,\mu} \alpha_{\nu,\mu}.
 \label{eq:overlap_manybody}
\end{align} 

To determine the MPO representation of $D^{O_1O_2}$ in Eq.~(\ref{eq:gate_coeffs}), we proceed as follows:  
\begin{enumerate}
\item[(i)] The Grassmann kernel $\mathcal{D}^{O_1O_2}$ in Eq.~(\ref{eq:expec_value_overlap}) naturally takes the form
\begin{equation}
\label{eq:kernel_factor_main}
\mathcal{D}^{O_1O_2} = \mathcal{D}_{M-1^+}\dots\mathcal{D}_0\dots \mathcal{D}_{M-1^-}.
\end{equation}
The Grassmann polynomial that is obtained by multiplying out the Grassmann kernels in Eq.~(\ref{eq:kernel_factor_main}) leads to Grassmann strings that are not ordered according to Eq.~(\ref{eq:GM_vec}). This is easy to see, since variables on the forward and backward branch with the same time index are placed next to each other in Eq.~(\ref{eq:GM_vec}), while they are separated in Eq.~(\ref{eq:kernel_factor_main}).
\item [(ii)] We permute the Grassmann variables to respect the ordering in Eq.~(\ref{eq:GM_vec}). This swapping of Grassmann variables generates non-trivial signs which are absorbed into the weights $\alpha_{\nu,\mu}$ in Eq.~(\ref{eq:kernel_factor_main}).
\item [(iii)] We include these signs {\it locally} in the kernels from Eq.~(\ref{eq:kernel_factor_main}), $\mathcal{D}_{m^\alpha} \xrightarrow{\text{signs}} \tilde{\mathcal{D}}_{m^\alpha},$ such that we can write
\begin{equation}\label{eq:GM_kernel_with_signs}
\mathcal{D}^{O_1O_2} = \Big(\tilde{\mathcal{D}}_{M-1^-}\cdot \tilde{\mathcal{D}}_{M-1^+}\Big) \dots \Big(\tilde{\mathcal{D}}_{0^-}\cdot \tilde{\mathcal{D}}_{0^+}\Big).
\end{equation} 
\item [(iv)]
Finally, we convert the \textit{Grassmann kernel} from each pair of brackets in Eq.~(\ref{eq:GM_kernel_with_signs}) to fermionic \textit{operators} by using Eqs.~(\ref{eq:coherent_state1}--\ref{eq:coherent_state2}), which amounts to replacing each Grassmann variable with a fermionic operator, similarly to how we rewrote the IF-kernel $\mathcal{I}$ as state $\ket{I}$ in Eq.~(\ref{eq:Istate}). This step allows us to define a product operator 
\begin{equation}\label{eq:product_operator}
D^{O_1O_2} = D_{M-1}\cdot D_{M-2}\dots D_1\cdot D_0,
\end{equation} where the individual operators $D_m$ can be viewed as effective tensors of a MPO with bond dimension $\chi=1$. Importantly, our procedure ensures that one finds precisely the weights $\alpha_{\nu,\mu}$ when explicitly evaluating the global many-body operator in Eq.~(\ref{eq:product_operator}).
\end{enumerate} 
 With this, Eq.~(\ref{eq:overlap_manybody})
 can be evaluated straight-forwardly as MPS-MPO-MPS contraction.  The full derivation of the kernels of the operators  $\tilde{\mathcal{D}}_{n}$ is detailed in App.~\ref{app:impurity_gate}. As mentioned in (iv), the kernels are directly converted into the fermionic operators resulting in the MPO tensors.

\subsubsection{Propagators}

The correlation functions of interest to us are 
the greater and lesser impurity Green's
functions, defined on the Keldysh contour with forward (superscript $+$) and backward (superscript $-$)
branch as \cite{Aoki2014}
\begin{eqnarray}
\label{eq:propagators}
G^>(t,t') &=& -i\langle d(t^-)d^\dagger(t'^+)\rangle_\text{imp}, \\ 
G^<(t,t') &=& i\langle d^\dagger(t'^-)d(t^+)\rangle_\text{imp}, 
\end{eqnarray}
which are analogous to Eq.~\eqref{eq_correlator}, 
with $O_1$ and $O_2$ corresponding to impurity creation or annihilation operators. 
The greater and
lesser components can be computed as shown in Fig.~\ref{IFMPS_Tensor_network}
and can be used to evaluate the retarded impurity Green's function
\begin{eqnarray}
G^R(t,t') &=& \theta(t-t')\lbrack G^>(t,t')  - G^<(t,t')\rbrack, 
\end{eqnarray}
from which the impurity spectral function
$A(\omega)=-\frac{1}{\pi}\mathrm{Im}G^R(\omega)$  can be calculated by linear interpolation between the discrete time-dependent data points and subsequent Fourier transformation.

\subsection{DMFT calculations}

\subsubsection{Model and lattice}

We will report DMFT results for the half-filled Hubbard model with Hamiltonian
\begin{equation}
H=-v_\text{nn}\sum_{\langle i,j\rangle}(d^\dagger_{i}d_j+\text{h.c.})+U\sum_i (n_{i\uparrow}-\tfrac12) (n_{i\downarrow}-\tfrac12),\label{eq_Hubbard}
\end{equation}
where the subscripts $i,j$ denote the lattice sites and the first sum is over nearest neighbor pairs. We consider 
an infinitely connected Bethe lattice, which has a semi-elliptical noninteracting DOS given by \cite{Georges1996} 
\begin{eqnarray}
A(\omega) = \frac{2}{\pi W}\sqrt{\left(\frac{W}{2}\right)^2 - \omega^2},
\label{Bethe_NI_hyb}
\end{eqnarray} 
where $W=4v$ is the bandwidth expressed in terms of the renormalized hopping
amplitude $v=\frac{v_\text{nn}}{\sqrt{Z}}$ ($Z\rightarrow \infty$ is the coordination number).
This scaling assures that for a generic value of the on-site repulsion $U$, the
kinetic and potential energies of the model are of the same order
\cite{Metzner1989}. 

DMFT maps the lattice problem \eqref{eq_Hubbard} to an impurity model of the type defined in Eqs.~\eqref{eqn_SIAM} and \eqref{eq_Himp}, with a self-consistently determined bath \cite{Georges1996}.

\subsubsection{Updating the hybridization function}
\label{sec_update_hyb}

A single iteration of the DMFT loop consists of computing the impurity Green's function, extracting the self-energy, calculating the lattice Green's function with this self-energy and imposing that the local lattice Green's function is the same as the impurity Green's function \cite{Georges1992}. In the case of an infinitely-connected Bethe lattice, the explicit calculation of the self-energy can be avoided, and the new hybridization function can be directly obtained from the impurity Green's function by the relation \cite{Georges1996,Aoki2014} 
\begin{equation}
\Delta(t,t') = v^*(t) G(t,t') v(t') = v^2 G(t,t'). \label{eq_bethe}
\end{equation}
The retarded component of this hybridization
function yields the specral function  
$A_{\mathrm{bath}}(\omega) = -\frac{1}{\pi}\mathrm{Im}\Delta^R(\omega)$, 
which is used for generating the $\mathbf{B}$ matrix
via Eq.~\eqref{Delta_defn}. With this updated $\mathbf{B}$ matrix,  the IF-MPS is recomputed. 

On a bipartite lattice, such as the Bethe lattice, the DMFT solution of the half-filled Hubbard model at high temperature crosses over from a paramagnetic metal ($U\lesssim W$) to a Mott-like bad metal ($U\gtrsim W$). At sufficiently low temperature, for $0<U<\infty$, an antiferromagnetic insulator is realized.
If antiferromagnetic order is suppressed, a metal-Mott insulator transition appears at low temperatures. 
In the paramagnetic phase, the IF-MPSs for up and
down spin are identical. To study the antiferromagnetic solution,  
one can combine the self-consistency relations for the two sub-lattices into the single relation \cite{Georges1996} 
\begin{eqnarray}
\Delta^\sigma(t,t') = v^2 G^{\bar\sigma}(t,t'). \label{eq_bethe2}
\end{eqnarray}
where $\bar\sigma$ stands for the spin opposite to $\sigma$. 
In an antiferromagnetic state, the spectral function becomes spin dependent and also the IF-MPS needs to be computed separately for the two spin species.

In equilibrium calculations, we use
$A_\text{bath}(\omega)=-\frac{1}{\pi}\text{Im}\Delta^R(\omega)$ in
Eq.~\eqref{Delta_defn} to define the new influence matrix for the given inverse temperature $\beta$. 
In the
nonequilibrium steady-state setup, we use a nonthermal distribution function
$f_\text{noneq}$ 
instead of the Fermi function $f_F$ in Eqs.~\eqref{nonint_les} and \eqref{nonint_gtr}. Specifically, to study photo-doped Mott states, we choose
a chemical potential $\mu_+$ ($\mu_-$) for the doubly occupied (empty) sites in
the upper (lower) Hubbard bands. 
The nonequilibrium distribution function is then defined through
correspondingly shifted Fermi-Dirac distributions with an
inverse effective temperature $\beta_\text{eff}$. 
Defining the half-frequency intervals $I_+=\omega\in[0,\infty)$ and 
$I_-=\omega\in(-\infty,0)$, the explicit formula reads 
\begin{equation}
f_\text{noneq}(\omega; \beta_\text{eff}, \mu_\pm)=\left\{
\begin{tabular}{ll}
$f_F(\omega; \beta_\text{eff},\mu_-)$, & $\omega \in I_-$\\
$f_F(\omega; \beta_\text{eff},\mu_+)$, & $\omega \in I_+$
\end{tabular}
\right. .
\end{equation}
In principle, one could employ a smooth switching function between the two half-intervals. However, if the gap in the spectral function is large enough, the details of the switching procedure do not matter much and we employ the above discontinuous switching. 

\subsubsection{Relaxation into a steady state}

As mentioned in Sec.~\ref{IF_formalism}, the calculation starts from a factorized density matrix [Eq.~\eqref{Eq_factorization}], so that the system exhibits a nontrivial time evolution while the entanglement between the impurity and the bath is built up. During this evolution, neither the equilibrium nor effective temperature are well-defined. After some relaxation time $t_\text{relax}$, the system reaches a time-translation invariant steady-state characterized by the imposed $\beta$ 
(equilibrium case) or $\beta_\text{eff}$ and $\mu_\pm$ (nonequilibrium case). One should measure the impurity Green's functions starting from a time larger than $t_\text{relax}$.

In practice, we use the diagonal components of the impurity density matrix at the maximum time $t=t_\text{max}$ as the initial impurity density matrix for the next iteration. However, the relaxation time into the steady state does not depend much on this initial choice. 

\subsubsection{Convergence}
\label{sec_convergence}
At the end of each DMFT iteration, one computes the spectral function from the retarded component of the propagators. The steady-state DMFT iterations are run until the absolute difference between two successive iteration's spectral functions decreases below $10^{-3}$ for all $\omega$ values. 
In simulations with low-temperature baths, the spectral functions can oscillate between successive DMFT iterations. In order to speed up convergence in such situations, we apply mixing, i.e., take the average of two successive iteration's spectral functions as the input for the next sweep.

Typically, the DMFT calculations need about 20 iterations for a given time step size when started from an input guess spectral function such as a semi-ellipse (metallic system) or two semi-ellipses (insulators). For smaller time step sizes, it helps to converge in smaller number of iterations by taking the converged solution from the bigger time step size DMFT iterations as the input. 

\begin{figure}
\centering
\includegraphics[height = 13.5cm, width = 13.5cm, keepaspectratio]{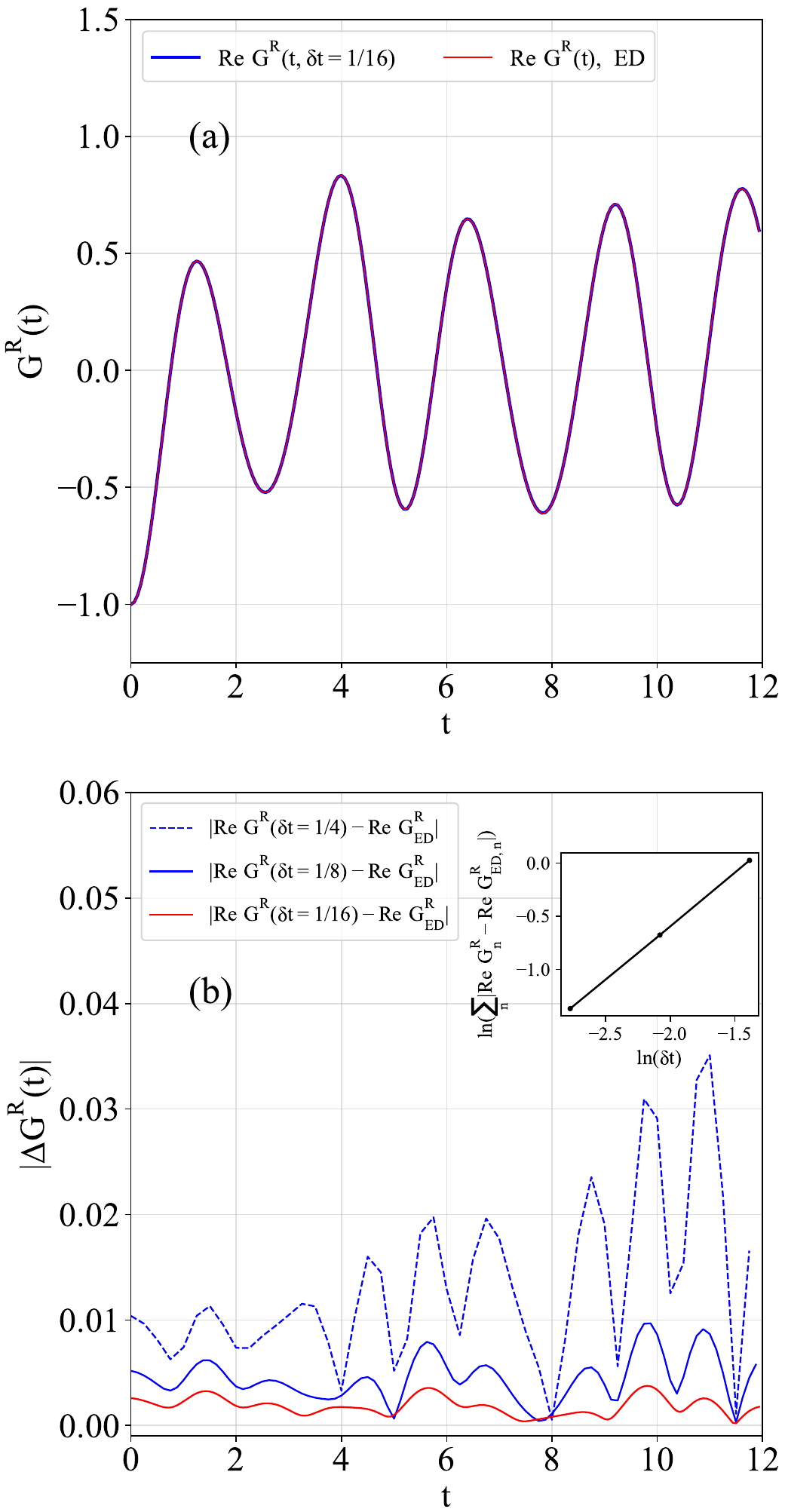}
  \caption{(a) Retarded Green's function for an impurity model with a single bath site ($V = 1, U = 4, T = 0.1$), computed using the IF-MPS method (blue) and exact diagonalization (ED, red). (b) Errors in the retarded Green's function when compared to the exact ED result, for the indicated values of the discretization step $\delta t$. Inset: A log-log plot with the $y$ axis corresponding to the sum over the time steps of the absolute differences between the retarded propagator and the ED propagator, and the $x$ axis given by the time step. 
  }
\label{IF_single_site_bath}
\end{figure}

\begin{figure*}
\centering
\includegraphics[height = 16cm, width = 16cm, keepaspectratio]{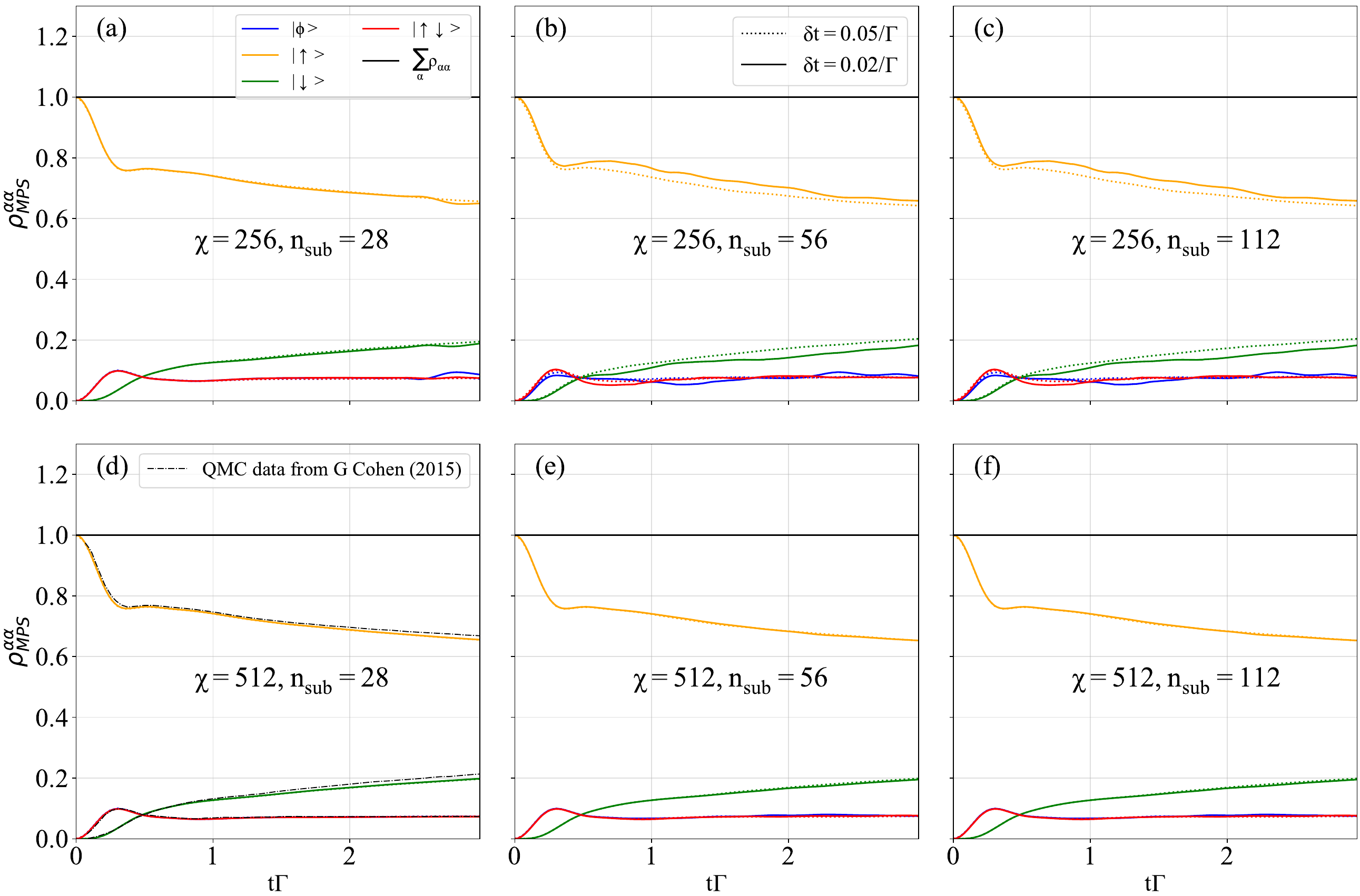}
\caption{Dependence of the diagonal elements of the impurity density matrix on the bond dimension, submatrix size and time step size, for the impurity model with wide rectangular bath density of states of height $\Gamma = 0.1$. The impurity interaction strength is $U = 8\Gamma$ and the initial configuration of the impurity density matrix is spin-up. The columns show the results for different sub-correlation matrix sizes $n_{\mathrm{sub}}$, and the two rows correspond to different bond dimensions $\chi$. The black dash-dotted lines in (d) show the inchworm quantum Monte Carlo (QMC) results from Fig.~2 of Ref.~\cite{Cohen2015}.}
\label{IF_rectangle_DOS_benchmarks}
\end{figure*}

\section{Impurity model results}
\label{sec_impurity}

Before employing the IF-MPS method described in the previous section as the impurity solver in DMFT calculations, we first show some benchmarks for impurity models with a fixed bath. 

In Fig.~\ref{IF_single_site_bath}, the impurity Green's function is computed for a model with a single bath site and hybridization amplitude $V$. The energy of the bath level is $\omega = 0$, the impurity level energy is $\epsilon_d=0$, the Hubbard on-site repulsion strength is $U = 4V$, the chemical potential is $\mu = 0$ and the temperature of the bath is  $T = 0.1V$. We measure energy in units of $V$ and time in units of $\hbar/V$ ($\hbar=1$).  For this simple model, one can compute the propagators using exact diagonalization, while the IF-MPS calculation based on the FW procedure is nontrivial. 
 The comparison between the exact result and the propagators computed for different finite step-sizes is shown in Fig.~\ref{IF_single_site_bath}(a). The almost perfect agreement shows that the MPS construction and the calculation of the correlation functions are implemented correctly. Fig.~\ref{IF_single_site_bath}(b) shows the difference between the IF-MPS data and the ED reference data for the different time step sizes. The error decreases proportionally to $\delta t$, consistent with a first-order scheme (see inset). We have also tested a second-order Trotter scheme (analogous to Ref.~\cite{Kloss2023}) instead of Eq.~\eqref{eq:Trotter}, but since the IF construction has an error $\mathcal{O}(\delta t)$, the overall error remains proportional to $\delta t$.

As a second benchmark, we consider an impurity model with a flat bath DOS with smooth edges, defined as 
\begin{eqnarray}
V^2A_{\mathrm{bath}}(\omega) = \frac{\Gamma}{\left(1+e^{\nu(\omega-\omega_c)}\right)\left(1+e^{-\nu(\omega+\omega_c)}\right)},
\end{eqnarray}
where $V=1$, $\omega_c = 10\Gamma$, $\nu = 10/\Gamma$ and $\Gamma=0.1$, as in Refs.~\cite{Cohen2015,Thoenniss2023a}. The temperature is set to $T = 0.02V$ and the chemical potential is set to $\mu = 0$. Figure~\ref{IF_rectangle_DOS_benchmarks} shows the evolution of the four impurity states in the fermionic basis (empty $|\phi\rangle$, singly occupied $|\sigma\rangle$, $\sigma=\uparrow,\downarrow$, and doubly occupied $|\!\uparrow\downarrow\rangle$), starting from the polarized initial density matrix $\rho_\text{imp}(t=0)=|\!\!\uparrow\rangle\langle \uparrow\!\!|$ with an impurity interaction strength of $U=8\Gamma$ at charge neutrality ($\epsilon_d=0$). The figure reports results for different time step sizes $\delta t$ as one varies the bond dimension $\chi$ (columns) and the sub-correlation matrix size $n_{\mathrm{sub}}$ (rows). We find that there is a breaking of particle-hole symmetry (different populations of empty and doubly occupied states) when $n_{\mathrm{sub}}\gtrsim 4\ln_2\chi$, 
which becomes increasingly severe for smaller $\delta t$. Hence, to avoid artifacts resulting from a large $n_\text{sub}$, the bond dimension $\chi$ must be chosen sufficiently large for the given time step size. In Fig. \ref{IF_rectangle_DOS_benchmarks}(d), which is roughly in line with the condition \eqref{eq:nsub_to_bond_dimension}, we compare the IF-MPS results for bond dimension $\chi =512$ and $n_{\mathrm{sub}} = 28$ with the inchworm QMC results from Fig.~2 of Ref.~\cite{Cohen2015} (black dashed curves) and find a good (although not perfect) agreement for the evolution of the local state populations. Since the IF-MPS calculation appears to be converged in the parameters of the IF formalism, we believe that our results are more accurate than the 10-year-old inchworm Monte Carlo data.

\section{DMFT results}
\label{sec_results}

\begin{figure*}
\centering
\includegraphics[height = 17cm, width = 17cm, keepaspectratio]{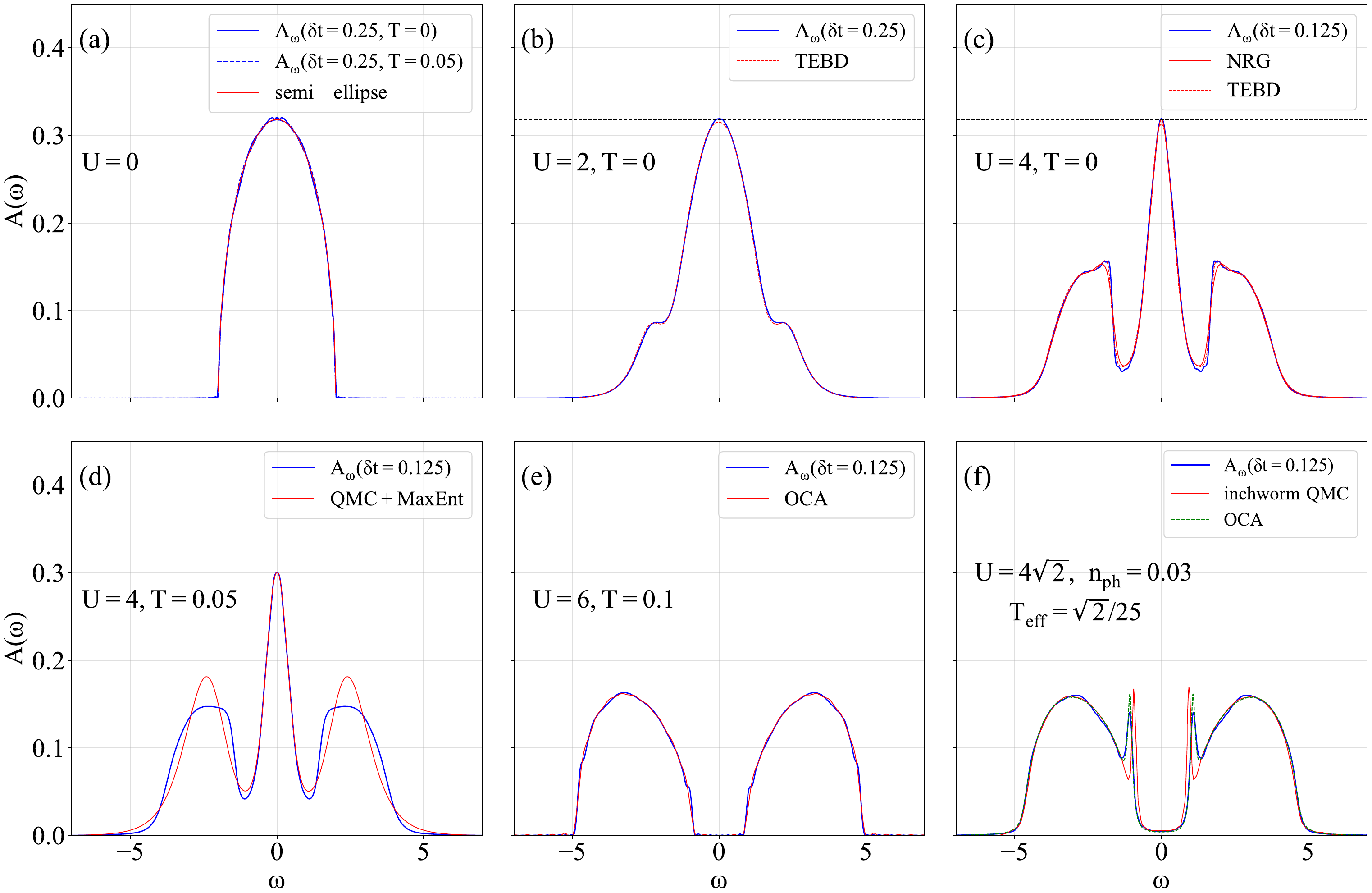}
  \caption{(a) Comparison of the IF-MPS spectral function for $U = 0$ with the exact semi-elliptical spectral function of the infinitely connected Bethe lattice. (b) Comparison of the IF-MPS spectral function for $U = 2$ and $T= 0$ with the TEBD data taken from Ref.~\cite{Ganahl2015}. (c) Comparison of the IF-MPS spectral function for $U = 4$ and $T=0$ with the TEBD data taken from Ref.~\cite{Ganahl2015} and  state-of-the-art NRG data (see text).
 The dashed lines in (b) and (c) show the expected value of the spectral function at $\omega=0$: $A(\omega = 0, T = 0) = 1/\pi$. (d) Comparison of the IF-MPS spectral function for $U = 4$ and $T = 0.05$ with the result obtained using continuous-time QMC and MaxEnt analytical continuation. (e) Comparison of the IF-MPS spectral function for $U = 6$, $T = 0.1$ with the result obtained from the one-crossing approximation (OCA). (f) Comparison of the IF-MPS spectrum for a photo-doped system with $U = 4\sqrt{2}$ and $T_\text{eff} =  \sqrt{2}/25$ with the inchworm QMC results from Ref.~\cite{Kuenzel2024} and the steady-state OCA spectrum \cite{Kim2024}.}
\label{IF_benchmarks}
\end{figure*}

\subsection{General remarks}

In this section, we report DMFT results for the half-filled Hubbard model on the infinitely connected Bethe lattice. First we compare IF-MPS spectra in the paramagnetic metallic and Mott insulating phase against results obtained by established methods to demonstrate the accuracy of the spectra. 
Then we show the effects of the discretization step $\delta t$ and of the product ansatz \eqref{Eq_factorization} for the initial density matrix. The challenges associated with a slow relaxation into the steady-state are illustrated by considering antiferromagnetic insulating systems with sharp spin-polaron peaks in the spectra. Finally, we demonstrate the suitability of the IF-MPS approach for the study of nonequilibrium steady-states by calculating spectral functions and occupation functions for photo-doped Mott insulators.

In this section, we use the renormalized hopping $v$ as the unit of energy and $\hbar/v$ as the unit of time ($\hbar=1$ in the following). In these units, the noninteracting bandwidth is $4$, the high-temperature end point of the Mott transition line in the paramagnetic phase is near $U=4.67$, $T=0.55$ and the zero-temperature end point is near $U\approx 5.8$ \cite{Bluemer2003}. In the case of an antiferromagnetic self-consistency loop, the Mott-transition line gets buried inside the antiferromagnetic insulating phase, and there merely exists a crossover from a weak-coupling to a strong-coupling antiferromagnet as $U$ is increased inside the insulating phase.

\subsection{Paramagnetic equilibrium states}

\subsubsection{Equilibrium benchmarks}

In Figs.~\ref{IF_benchmarks}(a)-\ref{IF_benchmarks}(e), we show benchmark calculations for the equilibrium Hubbard model in the paramagnetic state, specifically for the noninteracting system, weakly and moderately correlated metallic states and for the Mott insulating phase. For $U = 0$, independent of temperature $T$, the measured spectral function $A(\omega)=-\frac{1}{\pi}\mathrm{Im}G^R(\omega)$ should reproduce the semi-elliptical DOS \eqref{Bethe_NI_hyb} of noninteracting fermions on the infinitely coordinated Bethe lattice. This is indeed the case, apart from small wiggles at $T=0$, as demonstrated in Fig.~\ref{IF_benchmarks}(a). The likely origin of the wiggles is the slow decay of the correlations in time at low temperatures and the correspondingly demanding memory time cutoff or $n_{\mathrm{sub}}$ (which enforces a large bond dimension $\chi$ according to Eq.~\eqref{eq:nsub_to_bond_dimension}). At $T = 0.05$, the decay of the correlations in time is faster, which results in a smoother spectral function. While the impurity time evolution operators become trivial in the noninteracting case, this calculation is a good test for the precision of the IF-MPS construction.   

Introducing a weak (moderate) repulsive Hubbard interaction $U=2$ $(4)$, we next show results for the metallic phase. First, in Figs.~\ref{IF_benchmarks}(b,c), we compare our $T=0$ spectral function to the data from Ref. \cite{Ganahl2015}, which employs a MPS based method where the MPS represents the full system's density matrix at a given time. It is based on the time-evolving block decimation (TEBD) approach and directly computes the time-dependent correlations by applying the impurity and bath evolution gates to the ground state determined by the density matrix renormalization group (DMRG). We find a good agreement with this method, even for the higher energy Hubbard band substructures. In Fig.~\ref{IF_benchmarks}(c), 
we also compare our $T=0$ spectral function to state-of-the-art numerical renormalization group (NRG) data \footnote{Seung-Sup Lee (private communication). Improvements in the NRG algorithm as described in the Ref.~\cite{Lee2017} result in a fairly accurate determination of the higher energy Hubbard bands, as well as a highly accurate low-energy quasi-particle peak.}. In this case, 
the quasi-particle peak matches the IF-MPS result, while the structures of the Hubbard bands near $\omega=\pm \frac{U}{2}$ are similar, but differ by a small margin. The description of the Hubbard bands is challenging for NRG, which uses a logarithmic frequency grid. 

The IF-MPS calculations shown in Fig.~\ref{IF_benchmarks}(c) are for $\delta t=0.125$ and the description of the quasi-particle peak should further improve with smaller time steps. One sanity check is the integral of the spectral function, $w_A=\int d\omega A(\omega)$, which for $\delta t=0.125$ is 0.995, while the exact spectral function should satisfy $w_A=1$ \footnote{We renormalize the spectra at each iteration.}. Another sanity check is the value of the spectral function at $\omega=0$, which for the zero temperature solutions should be $A(\omega=0)=1/\pi$ (Friedel sum rule) \cite{Luttinger1961}. This condition is very well satisfied for the MPS-IF spectrum for $U=2$, while a small discrepancy is seen for $U=4$. At low frequencies, the IF-MPS spectrum is more accurate than the TEBD result, but not as accurate as state-of-the-art NRG. 

Figure~\ref{IF_benchmarks}(d) compares the IF-MPS spectral function for $T=0.05$ to the result obtained from a continuous-time quantum Monte Carlo (QMC) impurity solver \cite{Werner2006} via maximum entropy (MaxEnt) analytical continuation \cite{Jarrell1996}. The agreement is very good for the quasi-particle peak, and the positions of the Hubbard bands are consistent, whereas the structure of the Hubbard bands differs. It is well known that analytical continuation has difficulties predicting structures a high energy. 

The benchmark of a Mott insulating solution at $U=6$ is shown in Fig. \ref{IF_benchmarks}(e). Here, we do not have an exact reference sprectrum, but compare to the approximate result from the one-crossing approximation (OCA) \cite{Pruschke1989}. This self-consistent second order expansion around the atomic limit overestimates correlation effects, but should produce good result deep in the Mott regime. (For the chosen parameter set, the first and second order expansions give very similar spectra.) Since our OCA calculation is implemented on the three-branch Kadanoff-Baym contour \cite{Eckstein2010nca}, there are no inaccuracies associated with analytical continuation. The IF-MPS spectrum is consistent with the OCA result, but the calculation is heavy with $n_{\mathrm{sub}} = 40 $ and a bond dimension of 1024, indicative of a long memory time in low-temperature Mott insulators. The wiggles in the Hubbard bands get smaller with increasing number of DMFT iterations, and the IF-MPS result approaches the OCA spectrum. Figure~\ref{IF_benchmarks}(e) shows the almost converged spectrum after 12 iterations (a single iteration with an MPS of size 1920 takes 50 hours on 20 processors). 

\begin{figure*}
\centering
\includegraphics[height = 13cm, width = 13cm, keepaspectratio]{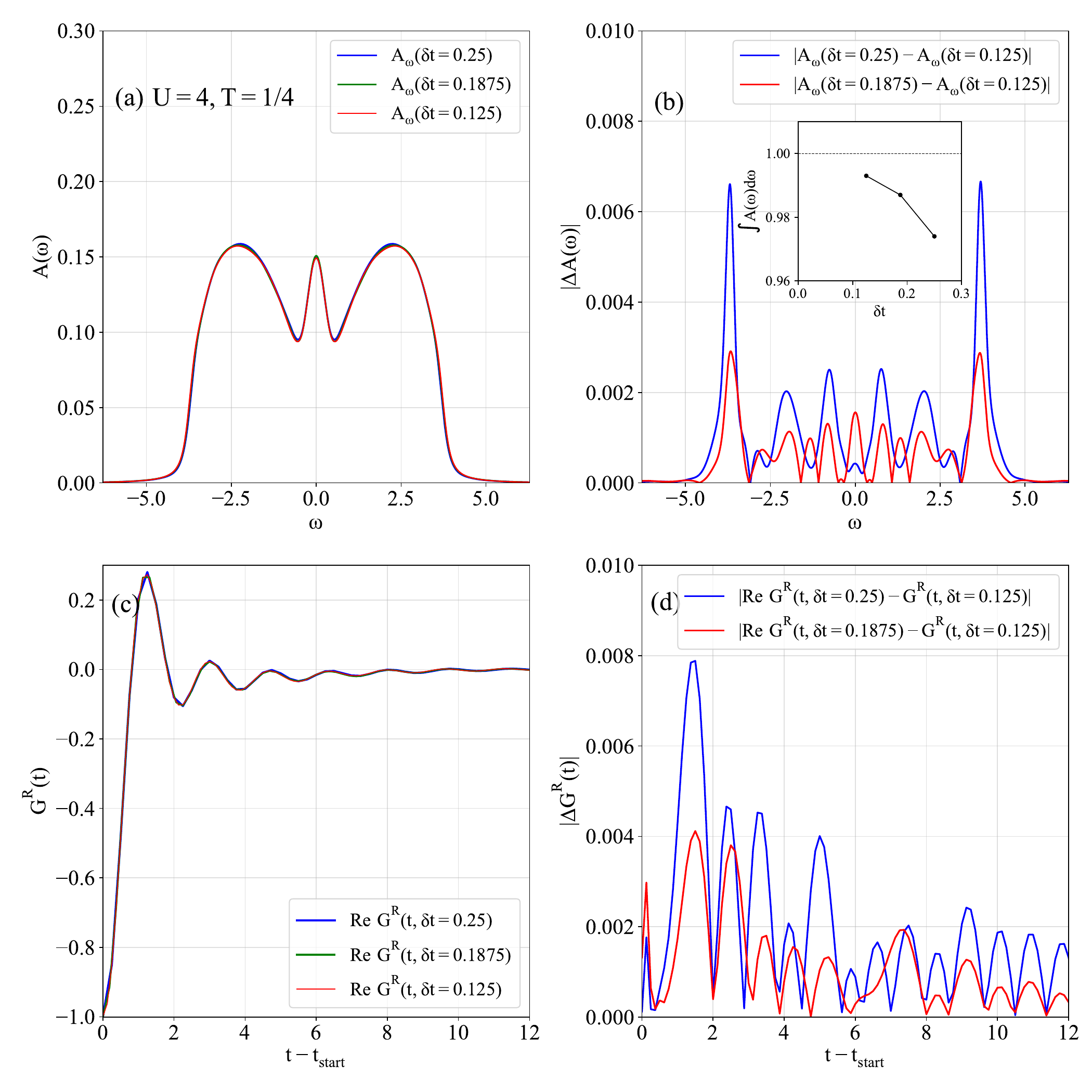}
\caption{Demonstration of the time step size dependence of the converged
  spectral function for $U = 4$ and $T = 0.25$. (a) $A(\omega)$,
  and (b) the absolute difference between the spectral functions for the
  indicated $\delta t$. The inset of (b) plots the spectral weight $w_A$ as a function of
  $\delta t$ and compares it to the exact result (horizontal dashed line). 
  (c) Real-time dependence of the retarded Green's
  function in the steady state for different $\delta t$. (d) Absolute
  difference between the retarded Green's functions for the indicated $\delta
  t$. The data in (a) and (c) for $\delta t = 0.25, 0.1875$ have
  been computed for bond dimension 512, while those 
  for $\delta t = 0.125$ are for bond dimension 1024.
}
\label{IF_step_size_dependency}
\end{figure*}

\subsubsection{Nonequilibrium benchmark}

In Fig.~\ref{IF_benchmarks}(f), we compare the IF-MPS nonequilibrium steady state solution for a photo-doped Mott insulator with photo-doping density $n_\text{ph}=0.03$, $U = 4\sqrt{2}$ and $T_\text{eff} =  \sqrt{2}/25$ 
to the inchworm QMC result from Ref.~\cite{Kuenzel2024}. The inchworm QMC scheme implements a high-order stochastic evaluation of the diagrams in the self-consistent strong-coupling expansion and works directly on the real-time axis (no need for analytical continuation). We find a good agreement in the shape of the Hubbard bands but the locations of the quasi-particle peaks associated with doublons (holons) are shifted to higher (lower) energy, compared to the inchworm spectrum. Furthermore, the height of the quasi-particle peaks is smaller. The origin of the discrepancies between these two ``numerically exact" spectra remains to be clarified. The IF-MPS spectrum is close to the OCA spectrum obtained by the recently developed TCI based steady-state approach~\cite{Kim2024}, which itself is close to the first-order (non-crossing approximation) result. This suggests that the strong coupling expansion converges fast for this setup, so that the exact spectrum should not be far from the OCA one. 

\begin{figure*}
\includegraphics[height = 17cm, width = 17cm, keepaspectratio]{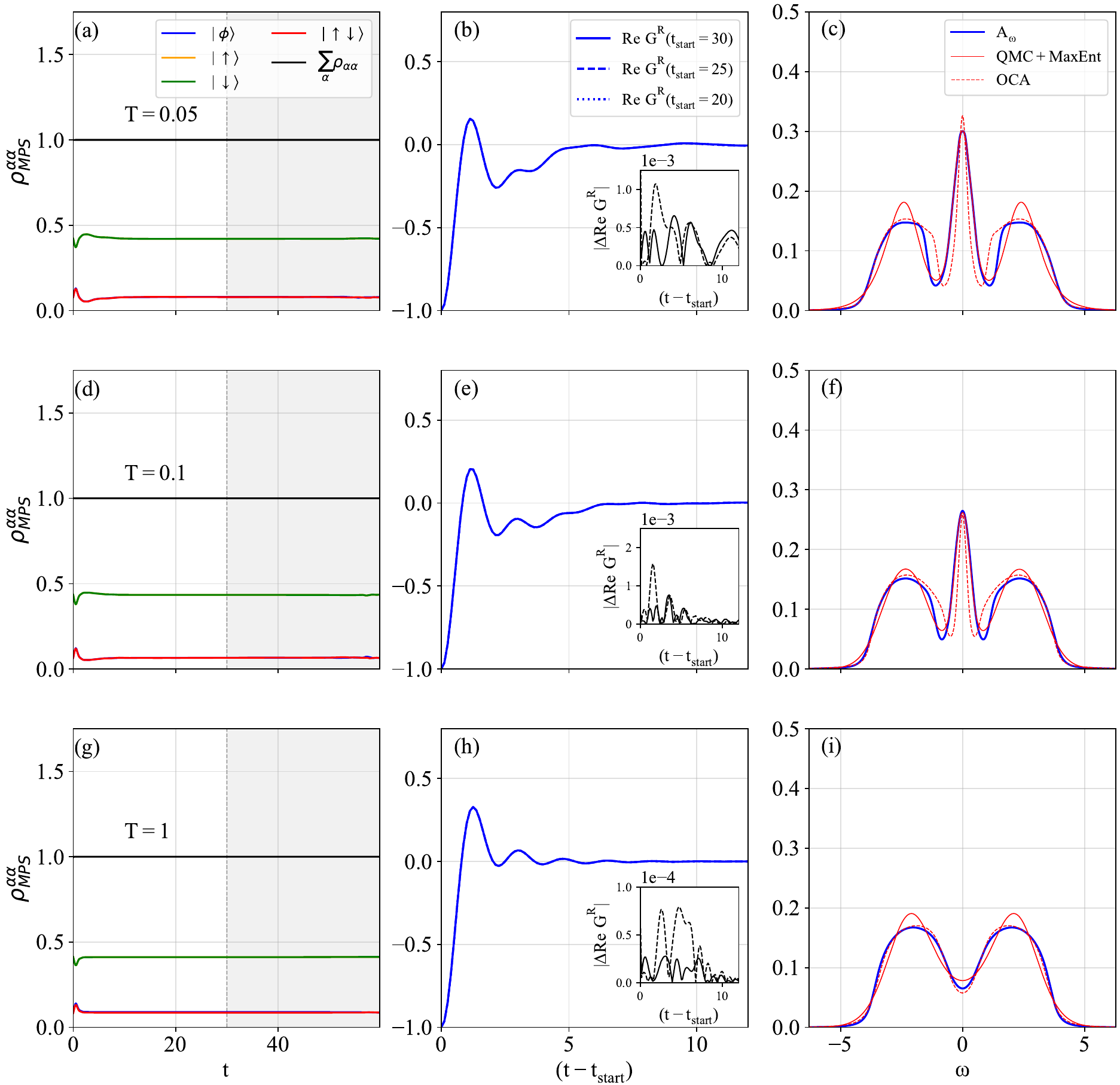}
\caption{Metallic DMFT solutions for  $U = 4$ and $T=0.05$ (top row), $0.1$ (middle row) and $1$ (bottom row), calculated for time step $\delta t  = 0.125$ and bond dimension $\chi = 1024$. In the left column (a), (d), (g), the evolution of the diagonal elements of the impurity density matrix is plotted. The dashed vertical line indicates $t_{\mathrm{start}}$ with $ t_{\mathrm{max}}$ fixed at 60. In the middle column (b), (e), (h), we plot the Green's function measured from $t_{\mathrm{start}}=20$, $25$ and $30$. 
The results lie on top of each other, so we plot in the insets the absolute difference between the Green's function for $t_{\mathrm{start}}=30$ and that for $t_{\mathrm{start}}= 25$ (solid black lines) and $t_{\mathrm{start}} = 20$ (dashed black lines). In the right column (c), (f), (i), the spectral functions from the IF-MPS method (blue solid lines) are compared to the QMC MaxEnt solutions (red solid lines) and OCA solutions (red dashed lines).}
\label{IF_metallic_solutions}
\end{figure*}

\subsubsection{Effect of the time step size}

The discretization of the influence functionals and time evolution operators results in discretization errors.  
In order to reach the final spectral function, we start with a time step size
$\delta t$ which is small enough that the full expected spectral bandwidth
is covered, i.e. $\Delta \omega \in (-\pi/2\delta t, \pi/2\delta t)$, and
converge the DMFT calculation. Then, we decrease the time step size, taking the
previously converged solution as the input and again run several DMFT
iterations. This procedure is repeated until the spectral function converges
(see Fig.~\ref{IF_step_size_dependency}(a,b)). As mentioned before, one measure for
the discretization error is the violation of the sum-rule for the spectral
function, $w_A = 1$. The inset of Fig.~\ref{IF_step_size_dependency}(b) shows how these deviations decrease with decreasing time step size.
For $\delta t=0.125$, we have less than 1\% deviation from the exact sum
rule. In practice, we normalize the spectra at each DMFT iteration. 

The convergence with decreasing time step size is also evident in the real-time
data for the retarded Green's function
(Fig.~\ref{IF_step_size_dependency}(c,d)). A short enough time step is needed
especially for capturing the fast initial decay of the Green's function and the
first oscillations. The details of the short-time decay matter for the
high-energy structures in the spectral function, while the damped oscillations
at longer times determine the quasi-particle peak. For the $U=4$ system with
relatively high temperature $T=0.25$ considered here, the quasi-particle peak is
suppressed and the Green's function is fully damped within a time window of
$\Delta t  \approx 16$. In a low-$T$ strongly correlated metal, or in an
antiferromagnetic state (Sec.~\ref{sec_afm}), sharp peaks in the spectral
function lead to a slower decay, so that a larger measurement window $\Delta t$
is required in the simulations.

\subsubsection{Relaxation into a steady state}

Since the IF-MPS simulations start from an impurity which is decoupled from the bath, one first needs to build up the entanglement with the bath and relax into a steady state corresponding to the desired (effective) temperature. The first row of Fig.~\ref{IF_metallic_solutions} illustrates this relaxation process for the system with $U=4$ and the indicated inverse temperatures. The full simulation covers a time interval of length $t_\text{max}=60$, and the four lines show the diagonal elements of the density matrix, i.e. the probabilities of the empty state ($|\phi\rangle$), the singly occupied states ($|\sigma\rangle$, $\sigma=\,\uparrow,\downarrow$) and the doubly occupied state ($|\!\!\uparrow\downarrow\rangle$). Since the system is paramagnetic and particle-hole symmetric, the weights of the spin-up and spin-down states, as well as the empty and doubly occupied states are equal.  

We start the solution with the density matrix measured at $t=t_\text{max}$ in the previous DMFT iteration. After the coupling to the bath is switched on at time $t=0_+$, the probabilities of the different local states change, but then relax back to the steady-state solution of the previous iteration after some time $t_\text{relax}$, which depends on the model parameters (here, $t_\text{relax}<20$ for all considered temperatures). 
Note, however, that the relaxation of the density matrix does not guarantee that the system has fully reached a steady state. Since the Green's function is the relevant observable in DMFT calculations, it is important to check that this function is indeed time translation invariant. We measure the impurity Green's function starting at a time $t_\text{start}$ which is larger than $t_\text{relax}$. The second column of Fig.~\ref{IF_metallic_solutions} plots the results for $t_\text{start}=20$, $25$ and $30$, which are essentially lying on top of each other. The insets show that the maximum deviations between the curves are of the order of $10^{-4}$-$10^{-3}$, which demonstrates that for these $t_\text{start}$, the measured Green's function is the steady-state result. 

Ideally, the Green's function fully decays within the measurement interval $[t_\text{start}, t_\text{max}]$, as in these examples. In the case of slowly damped oscillations, one can use linear prediction \cite{Levinson1946} to extrapolate the solution to longer times. This reduces cutoff-related artifacts in the spectra.

\subsubsection{Comparison of the spectra}

The third column of Fig.~\ref{IF_metallic_solutions} compares the IF-MPS spectra to the approximate spectral functions obtained with the one-crossing approximation and using MaxEnt analytical continuation of continuous-time QMC data \cite{Werner2006}. In the chosen parameter regime, with a not too narrow or completely suppressed quasi-particle peak, the IF-MPS spectrum can be regarded as numerically exact. The OCA solution provides a qualitatively correct description of these correlated metal states, but underestimates the weight of the quasi-particle peak and overestimates the width of the Hubbard bands. Analytical continuation of numerically exact Matsubara axis QMC data is expected to produce a good estimate of the quasi-particle peak at low temperatures, while high energy features and high-temperature spectra can only be qualitatively described. Indeed, the quasi-particle peaks of the QMC spectra in Figs.~\ref{IF_metallic_solutions}(c) and ~\ref{IF_metallic_solutions}(f) agree well with the IF-MPS results, while the detailed shape of the Hubbard bands cannot be captured by MaxEnt. Also the high-temperature incoherent metal spectrum is not accurately described by the QMC+MaxEnt approach. In contrast, the IF-MPS solver handles such systems efficiently, due to the short memory time. 

\subsubsection{Self-energy}
Besides spectral functions, the IF-MPS approach also provides reliable self-energies. The self-energy describes the effect of correlations on the propagation of the electrons. 
From the Dyson equation for the retarded component of the Green's function we get \cite{Aoki2014} 
\begin{eqnarray}
\Sigma^R(\omega) = \omega  
- \Delta^R(\omega) - \frac{1}{G^R(\omega)},
\end{eqnarray}
where $G^R(\omega)$ [$\Delta^R(\omega)$] is the Fourier transform of the retarded Green's function [hybridization function]. 
With the self-consistency relation \eqref{eq_bethe} for paramagnetic states, the above equation becomes
\begin{eqnarray}
\Sigma^R(\omega) = \omega -v^2 G^R(\omega) -\frac{1}{G^R(\omega)},\label{eq_self}
\end{eqnarray}
where $v=1$ in our convention.  
\begin{figure*}
\centering
\includegraphics[height = 15cm, width = 15cm, keepaspectratio]{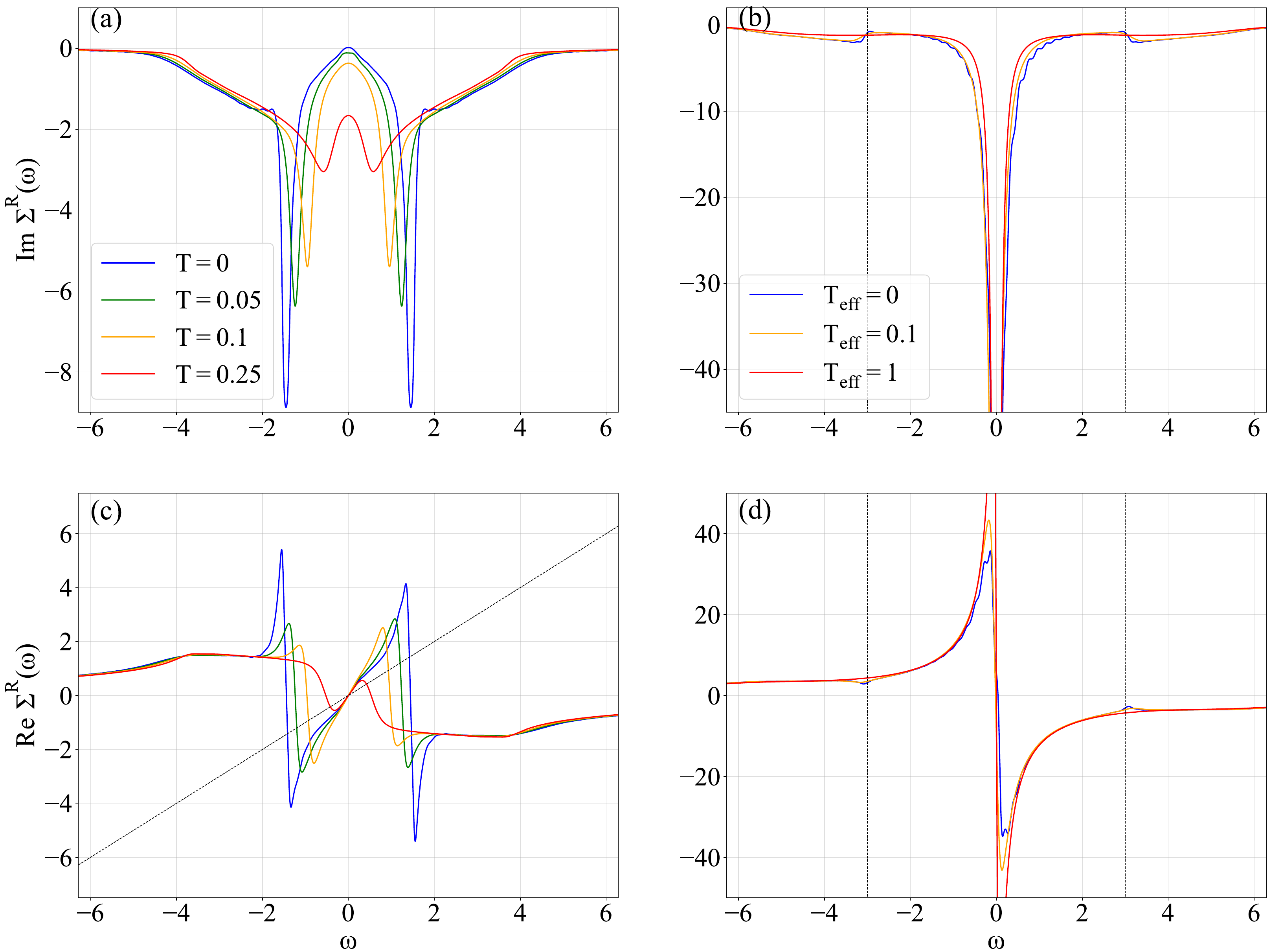}
\caption{Self-energy $\Sigma^R(\omega)$ as a function of frequency, computed at different temperatures using $\delta t = 0.125$ and bond dimension $\chi = 1024$. The upper and lower row shows the imaginary and real part of $\Sigma^R(\omega)$, respectively. (a), (c) Equilibrium system with $U=4.$ The black dashed line corresponds to $y=\omega$. (b), (d) Photo-doped system with $U=8, \mu_{\pm}=\pm 3.$ The black vertical lines indicate the effective chemical potentials of the doublons and holons ($\mu_{\pm} = \pm3$).}
\label{self_energy_figure}
\end{figure*}
In Fig.~\ref{self_energy_figure}, we show the results obtained for the equilibrium Hubbard model with $U = 4$. In Fig.~\ref{self_energy_figure}(a), the imaginary part of the retarded self energy is plotted for different temperatures. 
Since the low-frequency value is proportional to the inverse lifetime of the quasi-particles, the data in this panel show an increasingly longer lifetime as temperature is decreased. At $T=0$, we find that the imaginary part of the self-energy vanishes quadratically near $\omega=0$, as expected for a Fermi liquid. 
This result is consistent with the $T=0$ NRG self-energy plotted in Ref.~\cite{Bulla1999}, except for discrepancies at higher $\omega$, which can be explained by the limitations of NRG. 
In Fig.~\ref{self_energy_figure}(c) we plot the corresponding real parts of the self energy. The intersection between the $y = \omega$ line (black dashed) with the $y=\mathrm{Re}\Sigma^R(\omega)$ line roughly locates the position of the quasi-particle peak and the Hubbard bands.

\subsubsection{Numerical effort}

Let us briefly comment on the computational cost of the different calculations. The simulations with continuous-time QMC \cite{Werner2006} on the Matsubara axis and the analytical continuation take between a few hours and one day (depending on the precision), on a single processor. The implementation of such calculations on the real-time axis comes with a sign problem \cite{Werner2009} and a computational effort that scales exponentially with $t_\text{max}$.  
The OCA calculations on a three-branch Kadanoff-Baym contour with $t_\mathrm{max}=21$ take about four days on a single processor and the cost increases like the fourth power of $t_\mathrm{max}$. 
(Here, shorter $t_\text{max}$ can be used, compared to the IF-MPS approach, because the initial state is entangled with the bath.) The recently developed TCI implementation of steady-state OCA brings the cost of OCA calculations down by at least two orders or magnitude \cite{Kim2024,Eckstein2024}.
The IF-MPS method takes 12 hours on 10 processors for one iteration with $\delta t = 0.25$, bond dimension $\chi=512$ and $t_\mathrm{max}=120$, and most of the computational time is spent on computing the IF-MPS. With fixed bond dimension, the calculation cost scales linearly with the length of the IF-MPS generated, i.e. $L_{\mathrm{MPS}} = 4t_{\mathrm{max}}/\delta t$, but in practice we found that the bond dimension has to be increased if the time step size is decreased ($\propto 1/\delta t$), making the effective cost proportional to $t_\mathrm{max}/{(\delta t)}^4$. 
The full steady-state DMFT calculation takes about 10 iterations at high temperatures and about 20 iterations at low temperatures. Hence, IF-MPS calculations are substantially more expensive than 
OCA or Matsubara-axis QMC calculations, but they also provide more accurate spectra. All the calculations shown in this paper can be obtained with small-scale parallelization.

\subsection{Antiferromagnetic equilibrium states}
\label{sec_afm}
\begin{figure*}
\centering
\includegraphics[height = 15cm, width = 15cm, keepaspectratio]{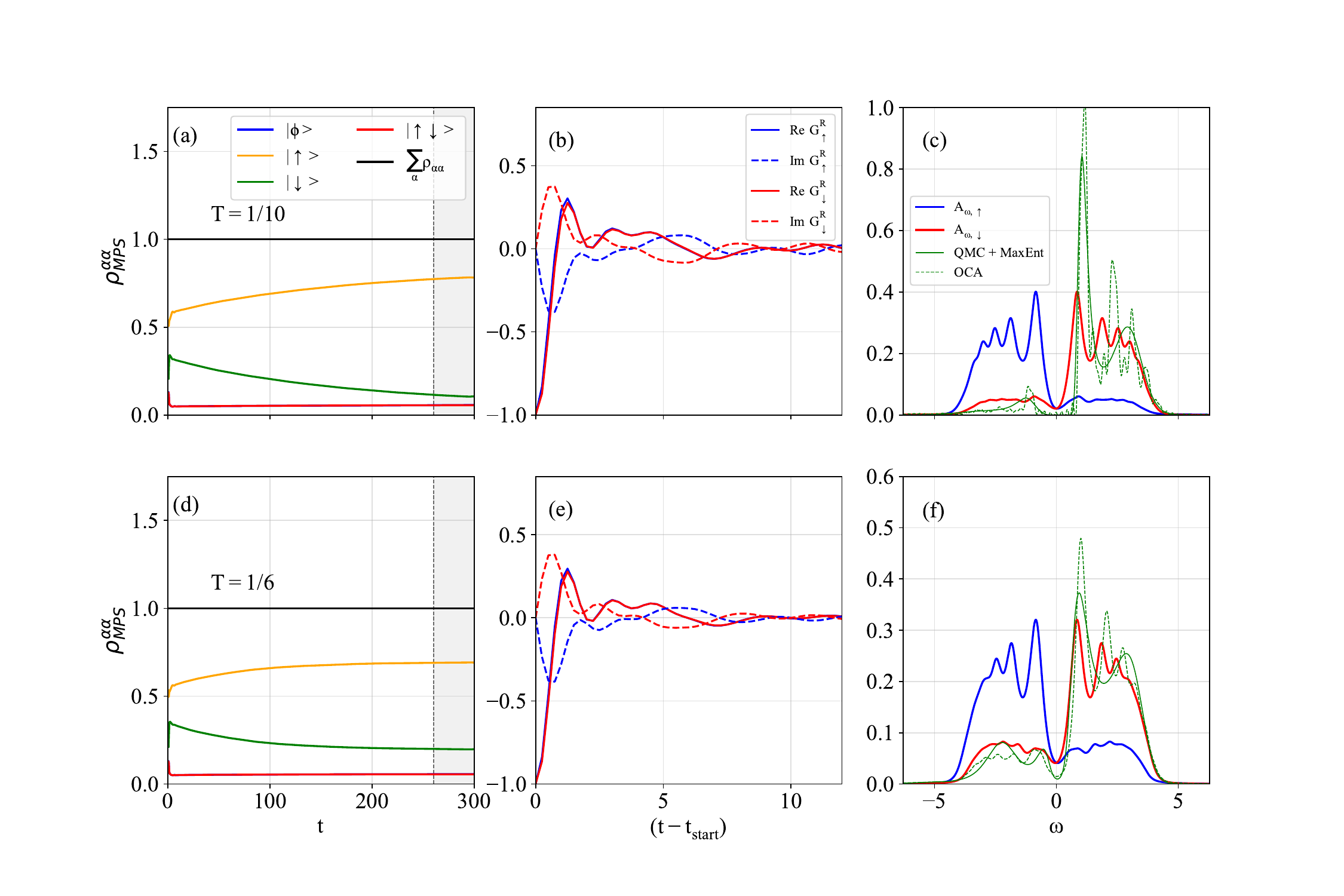} 
\caption{Antiferromagnetically ordered solutions for the half-filled Hubbard model with $U = 4$ and $T=0.1$ (top row), $0.167$ (bottom row). The simulations are for $\delta t  = 0.25$, bond dimension 512 and $t_{\mathrm{max}}$ up to $300$. In the left column (a), (d), the evolution of the diagonal elements of the impurity density matrix are plotted. In the middle column (b), (e), we plot the measured Green's functions for up-spin (blue) and down-spin (red). In the right column (c), (f), the spectral functions computed with the IF-MPS method (shown in blue and red for up and down-spin, respectively) are compared to the down-spin spectral functions computed using continuous-time QMC and MaxEnt (solid green) and OCA (dashed green).}
\label{AFM_solutions}
\end{figure*} 

The DMFT solution of the half-filled Hubbard model is antiferromagnetically ordered at low $T$. To describe this state, we introduce spin-dependent Green's functions $G^\sigma$ and hybridization functions $\Delta^\sigma$, and use the antiferromagnetic self-consistency condition \eqref{eq_bethe2}.

In the antiferromagnetic state, the majority-spin (minority-spin) spectral function gains (loses) occupation and the Hubbard bands split up into spin-polaron bands. At low temperature these side-band features become sharp, resulting in a slowly decaying Green's function. Also the build-up of the entanglement between the impurity and the bath becomes noticeably slower than in the paramagnetic state \footnote{In Sec. \ref{sec_impurity}, the evolution of the population densities of individual spin species is obtained by tracking the diagonal elements of the effective impurity density matrix at each time step. This is same as computing the expectation value of population density operators corresponding to different spin species with the IF-MPS . The population densities computed directly from expectation value is presented in the Fig. \ref{AFM_solutions} and shows a very slow decay into steady state.}. 
The interaction $U=4$ places our system in the intermediate coupling region with maximum N\'eel temperature ($T_\text{N}\approx 0.2$) \cite{Werner2012}. 
 
Because of the slow relaxation into a time-translation invariant state, we use a larger timestep $\delta t=0.25$ in these calculations. With this, we can push the measurement time of the Green's function to $t_\text{start}=260$, which however is still not enough to reach the steady state at $T=0.1$. This presumably explains why the IF-MPS calculation underestimates the magnetization (asymmetry in the up and down spin spectra), as well as the gap in the spectral function. In Fig.~\ref{AFM_solutions}(b), we compare the minority spin IF-MPS spectral function (red curve) to the spectra obtained from OCA (green dashed lines) and from continuous-time QMC via analytical continuation (solid green lines). The MaxEnt spectrum cannot be expected to capture the substructures of the Hubbard bands, but it should give a reliable estimate of the gap and roughly reproduce the first spin-polaron peak. Indeed, these two features match well between the MaxEnt and OCA results. This strongly suggests that the height of the spin-polaron peaks and the gap size are underestimated in the IF-MPS spectra. Consistent with this conclusion, we observed that both grow if we increase $t_\text{start}$. The same is true for the higher-temperature spectra shown in Fig.~\ref{AFM_solutions}(d), even though in this case, the density matrix has reached a steady state within the accessible time window. The shown IF-MPS spectra are thus not for the true equilibrium state at the indicated temperatures, but still influenced by the slow relaxation from the initial product state. They are representative of what can be obtained with moderate computational effort (by generating two IF-MPS each of length 4800 and bond dimension 512 in each DMFT iteration) using our current implementation.  

In the exact solution for the half-filled system, the real parts of the up and down Green's function should be identical, while the imaginary parts should have opposite signs. 
As shown in Fig.~\ref{AFM_solutions}(a,c), this is indeed the case, up to a small artificial symmetry breaking. 
We attribute the latter to the time step $\delta t=0.25$, which is too large for accurate low-temperature calculations. 
 
 \begin{figure}[t]
\centering
\includegraphics[height = 8.5cm, width = 8.5cm, keepaspectratio]{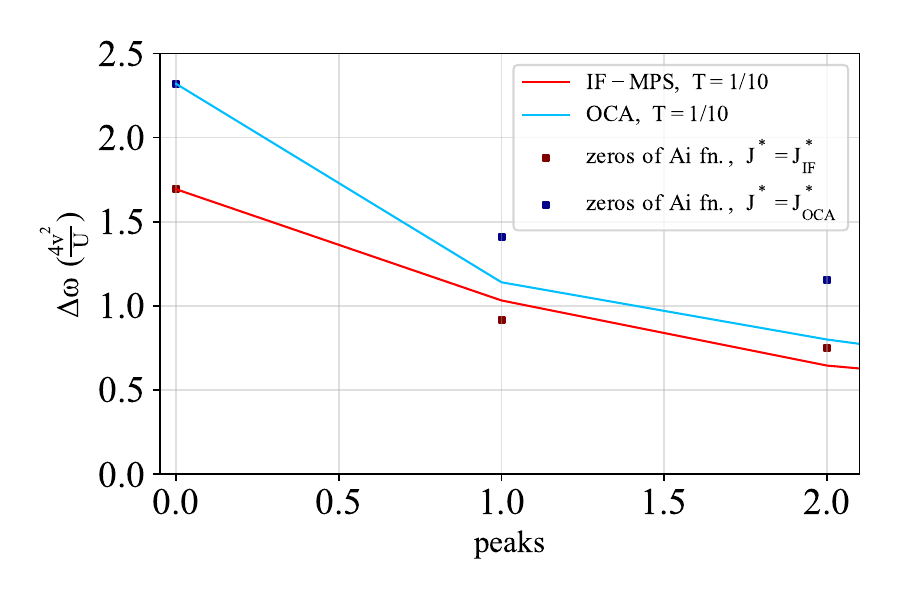}
\caption{Energy difference between successive spin-polaron peaks in the antiferromagnetic solution of the half-filled Hubbard model with $U = 4$ and $T=0.1$. The square symbols are the difference between the expected energy positions for a single hole in the $t$-$J$ model, where $J$ is selected such that the first spin-polaron peak position in the analytical expression matches the numerically obtained spectrum. $J^*_{\mathrm{IF}}$ denotes the renormalized value for the IF-MPS solution, while $J^*_{\mathrm{OCA}}$ denotes the renormalized value for the OCA solution (see text).}
\label{AFM_peaks_difference}
\end{figure}

The splittings between the spin-polaron peaks are smaller in the IF-MPS spectra, compared to the OCA results. For the $T=0.1$ spectrum, we plot these splittings  
in Fig. \ref{AFM_peaks_difference}. The spin polaron peaks of the lower (upper) Hubbard band describe holons (doublons) dressed by a spin cloud. 
The motion of a hole in the N\'eel ordered spin configuration leads to a string of misaligned spins, which results 
in a linear confining potential. The energy positions of the spin-polaron peaks can be analytically calculated for a single hole in the $t$-$J$ model \cite{Strack1992}. Relative to the innermost edge of the semi-elliptic Hubbard band, the peaks are predicted at
\begin{eqnarray}
\label{eq:Airy_fn_energies}
E_n = -\frac{J^*}{2} - a_n v\left(\frac{J^*}{2v}\right)^{\frac{2}{3}}, 
\end{eqnarray}
where $J^*$ is the effective spin-exchange interaction  and $a_n$ is the $n$-th zero of the Airy function. To compare these analytical results to our Hubbard model spectra, we tune the $J^*$ value such that the position of the first spin polaron peak aligns with Eq.~\eqref{eq:Airy_fn_energies}, which gives  $J^*_{\mathrm{IF}} \sim 0.76$ in units of $4v^2/U$. The location of the second predicted peak lies slightly lower in energy to that of the IF-MPS spectrum, while the third spin polaron peak is closer to the predicted energy levels in Eq.~\eqref{eq:Airy_fn_energies}. As a result the difference between the peak energies measured in IF-MPS spectrum compare well with the predicted energy differences. If we apply the same procedure to the OCA spectrum ($J^*_{\mathrm{OCA}} \sim 1.444$ in units of $4v^2/U$), we find deviations to the analytical prediction already for the second spin-polaron peak.

\subsection{Photo doped Mott system}

We now switch to the simulation of nonequilibrium steady states in Mott insulating Hubbard models. Photo-excitation of charge carriers across the Mott gap creates doublons (doubly occupied sites) and holons (empty sites) whose life-time grows exponentially with $U$ \cite{Sensarma2010}. If these charge carriers couple to some environment, they can dissipate their kinetic energy and the system may reach a long-lived state with a distribution of doublons (holons) in the upper (lower) Hubbard band described by an effective temperature $T_\text{eff}$ \cite{Murakami2023}. Such a state can for example be stabilized by weakly coupling the Hubbard bands to electron baths within the nonequilibrium steady state formalism \cite{Li2021}. Recently, it was shown that one can also directly impose the nonequilibrium distribution corresponding to $T_\text{eff}$ and converge to a self-consistent solution for the retarded and lesser Green's function \cite{Kuenzel2024}. 

\begin{figure*}
\includegraphics[height = 17cm, width = 17cm, keepaspectratio]{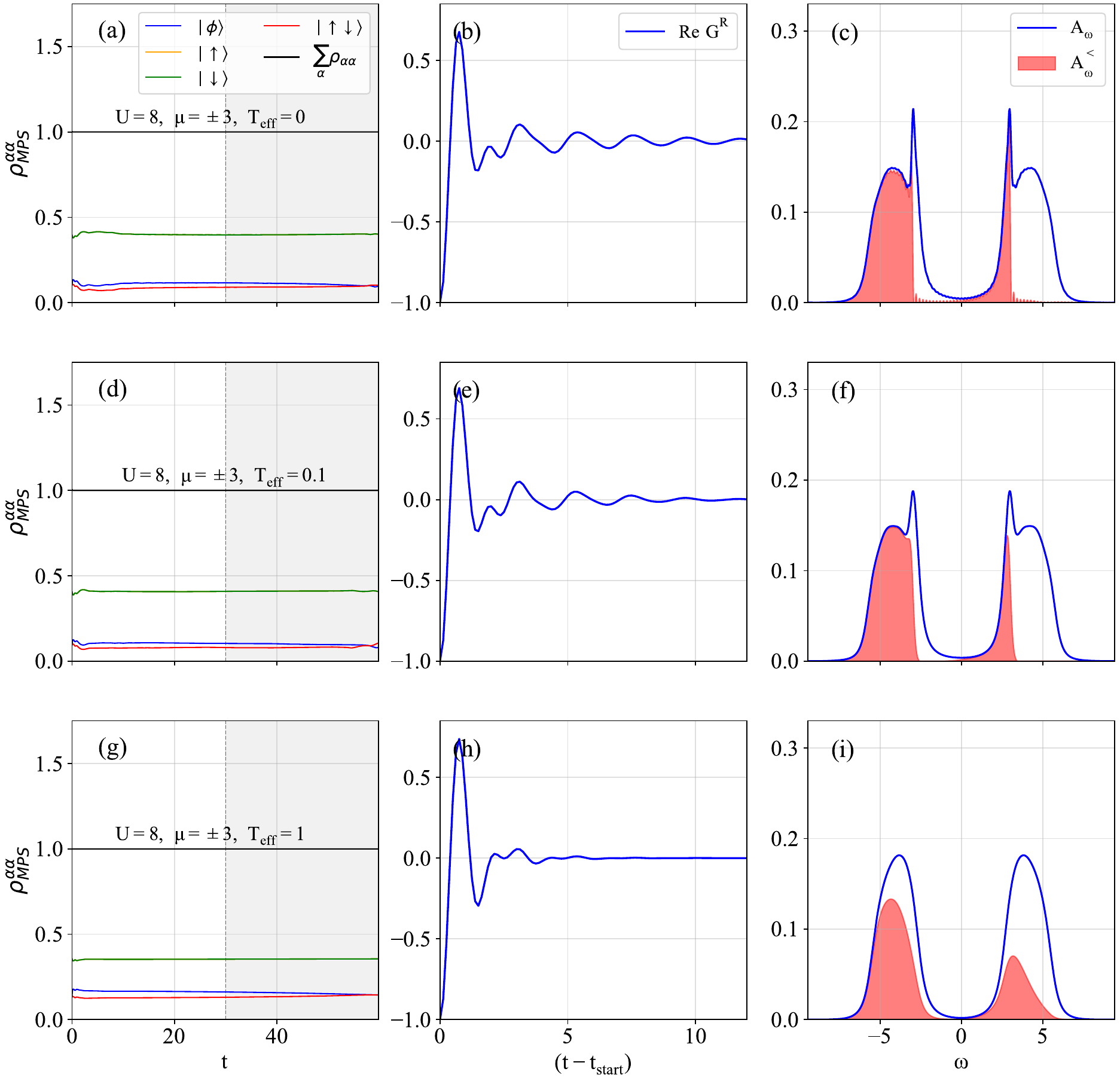}
\caption{DMFT solutions for photo-doped nonequilibrium steady states of the Hubbard model with $U = 8$ and different $T_\text{eff}$. The nonequilibrium state is produced by setting $\mu_\pm = \pm 3$ and we use time step sizes $\delta t  = 0.125$ (with bond dimension 1024). In the left column (a), (d), (g), the diagonal elements of the impurity density matrix are plotted. The doublon and holon weights are identical at $t=0$, but an artificial symmetry breaking occurs at $t>0$. The propagators were measured in the region shaded in grey. The middle column (b), (e), (h) plots the measured Green's functions. 
In the right column (c), (f), (i), the spectral functions computed with the IF-MPS method are shown by the blue lines. The boundary of the red shaded region corresponds to the computed occupation ($A^<(\omega)$).}
\label{Photodoped_solutions}
\end{figure*}

Here, we choose the latter approach and impose a nonequilibrium distribution $f_\text{noneq}(\omega)$ with two chemical potentials at $\mu_\pm$, as described in Sec.~\ref{sec_update_hyb}. The results for $U=8$, $\mu_\pm=3$ and different $T_\text{eff}$ are shown in Fig.~\ref{Photodoped_solutions}. The blue lines in the third column show the spectral function $A(\omega)=-\frac{1}{\pi}\text{Im}G^R(\omega)$ and the red shading the measured occupation $A^<(\omega)=-\frac{1}{2\pi}\text{Im}G^<(\omega)$. The latter satisfies the condition $A^<(\omega)=f_\text{noneq}(\omega)A(\omega)$ for a self-consistent solution. In the result for $T_\text{eff}=0$, we notice however small oscillations, which indicates that some simulation parameters ($\delta t$, $\chi$, or $n_\text{sub}$) are not yet optimally chosen. 
Overall, however, the low-$T_\text{eff}$ spectra show the expected features of an effectively cold photo-doped state, namely two quasi-particle peaks at the edges of the Hubbard bands, associated with the mobile doublons and holons \cite{Eckstein2013}. At the highest effective temperature, $T_\text{eff}=1$, these quasi-particle peaks disappear and the spectra show two featureless Hubbard bands with a broad doublon and holon distribution. The self-energies for these solutions, calculated using Eq.~\eqref{eq_self}, are plotted in Fig.~\ref{self_energy_figure}(b,d).

The left and middle columns of Fig.~\ref{Photodoped_solutions} show the time evolution of the diagonal elements of the density matrix and the retarded Green's function, respectively. In the density matrix results we notice an anomaly in the evolution of the doublon and holon probabilities. These should be degenerate in our particle-hole symmetric system, but the simulation somehow breaks this degeneracy. 
Since we explicitly symmetrize the spectral functions obtained at each iteration, this symmetry breaking must happen in the construction of the influence functional, and it is likely related to the 
way in which we initialize the FW algorithm. Decreasing the time step and increasing the bond dimension decreases this artificial splitting, but the issue persists for $\delta t=0.125$ and $\chi=1024$, which is already a computationally heavy calculation. Alternative procedures for computing the IF-MPS \cite{sonner2025semi} will hopefully resolve this issue.  

\section{Conclusions}
\label{sec_conclusions}
We applied the recently developed IF-MPS approach as impurity solver in DMFT calculations, and computed real-frequency spectral functions of the Hubbard Model on a Bethe lattice with infinite coordination. 
We demonstrated that in equilibrium settings the IF-MPS method can match or even surpass 
the accuracy of established methods, at least in the weakly to moderately correlated metal regime, where the IF-MPS approach works well independent of temperature. 
With the current implementation, it is however not yet possible to efficiently simulate low-temperature strongly correlated metals or antiferromagnetic states with sharp quasi-particle or spin-polaron peaks. The method also struggles to produce fully converged solutions for low-temperature Mott insulators with large gap. 
For observables which can be computed on the Matsubara axis, like the magnetization of equilibrium systems, the IF-MPS method as implemented here is not competitive with QMC solvers.

The influence functional approach is a particularly promising impurity solver for nonequilibrium steady-state simulations, where it enables for example an accurate description of photo-doped Mott states. For such applications, only few alternative numerically exact methods such as inchworm Monte Carlo \cite{Cohen2015,Kuenzel2024} exist. Also, in this context, one is not necessarily interested in very low effective temperatures, which are hard to reach in experiments. For not too low $T_\text{eff}$, the steady-state IF-MPS solver can provide numerically exact nonequilibrium DMFT results with small scale parallelization.
 
While the calculations presented in this work have been run on a computer cluster of modest size, the simulations are still too expensive for widespread applications. The main bottleneck is the calculation of the IF using the FW algorithm. If this step can be significantly improved, for example using the recently proposed semigroup influence matrix method \cite{sonner2025semi}, or using TCI \cite{Fernandez2022}, the IF-MPS solver will become practically useful for a broad range of DMFT applications. In the strongly correlated regime, high-order strong-coupling impurity solvers based on TCI \cite{Fernandez2022,Eckstein2024,Kim2024} will likely outperform the IF-MPS solver, but for moderately correlated metallic systems  and low-temperature states, the IF approach should be competitive.

\acknowledgements

We thank Aaram Kim for providing the steady-state OCA spectrum shown in Fig.~\ref{IF_benchmarks}(f) and Seung-Sup Lee for the NRG spectrum shown in Fig.~\ref{IF_benchmarks}(c). We also thank Lei Wang, Chu Guo and Antoine Georges and Jan von Delft for helpful discussions.
The calculations were run on the beo06 cluster at the University of Fribourg.
MN and PW acknowledge support from SNSF Grant No.~200021-196966. This work was also partially supported by the European Research Council via Grant Agreement TANQ 864597 (JT and DA).
\section*{Data Availablity}
The data that support the findings of this article are openly available \cite{nayak_2025_15611163}.

\appendix

\section{Justification of Eq.~\eqref{Green's_function_matrix_defn1}}
\label{app_A}
Introducing the spinors $\Psi_{\sigma,n}=\begin{pmatrix}
d^\dagger_{\sigma,{2n}^+}
&d_{\sigma,{2n}^-}
&d_{\sigma,{2n+1}^+}
&d^\dagger_{\sigma,{2n+1}^-}
\end{pmatrix}$
the Green's function matrix expressed in the time-dependent fermionic basis is given by
\begin{widetext}
\begin{multline}
{\bm{\widetilde G}}_{\sigma,m>n}= \langle \Psi_\sigma^T \Psi_\sigma \rangle_\text{imp}\Bigg|_{U=0, \epsilon_d = 0} \\
=  \frac{1}{2}\begin{pmatrix}
\langle d^\dagger_{\sigma,{2m}^+}d^\dagger_{\sigma,{2n}^+}\rangle_\text{imp} & \langle d^\dagger_{\sigma,{2m}^+}d_{\sigma,{2n}^-}\rangle_\text{imp} & \langle d^\dagger_{\sigma,{2m}^+}d_{\sigma,{2n+1}^+}\rangle_\text{imp} & \langle d^\dagger_{\sigma,{2m}^+}d^\dagger_{\sigma,{2n+1}^-}\rangle_\text{imp}\\
\langle d_{\sigma,{2m}^-}d^\dagger_{\sigma,{2n}^+}\rangle_\text{imp} &\langle d_{\sigma,{2m}^-}d_{\sigma,{2n}^-}\rangle_\text{imp} & \langle d_{\sigma,{2m}^-}d_{\sigma,{2n+1}^+}\rangle_\text{imp} & \langle d_{\sigma,{2m}^-}d^\dagger_{\sigma,{2n+1}^-}\rangle_\text{imp}\\
\langle d_{\sigma,{2m+1}^+}d^\dagger_{\sigma,{2n}^+}\rangle_\text{imp}  & \langle d_{\sigma,{2m+1}^+}d_{\sigma,{2n}^-}\rangle_\text{imp} & \langle d_{\sigma,{2m+1}^+}d_{\sigma,{2n+1}^+}\rangle_\text{imp} & \langle d_{\sigma,{2m+1}^+}d^\dagger_{\sigma,{2n+1}^-}\rangle_\text{imp}\\
\langle d^\dagger_{\sigma,{2m+1}^-}d^\dagger_{\sigma,{2n}^+}\rangle_\text{imp}  & \langle d^\dagger_{\sigma,{2m+1}^-}d_{\sigma,{2n}^-}\rangle_\text{imp} & \langle d^\dagger_{\sigma,{2m+1}^-}d_{\sigma,{2n+1}^+}\rangle_\text{imp} & \langle d^\dagger_{\sigma,{2m+1}^-}d^\dagger_{\sigma,{2n+1}^-}\rangle_\text{imp}
\end{pmatrix}\Bigg|_{U=0, \epsilon_d = 0},
\end{multline}
\end{widetext}
where the expectation values are for a noninteracting level with energy $\omega$. Since the corresponding Hamiltonian is number-conserving, the expectation values with two annihilation or two creation operators are zero. The other expectation values correspond to the lesser and greater Green's functions \eqref{nonint_les} and \eqref{nonint_gtr} of the noninteracting level (here dropping the spin label as the following expressions hold for each IF individually):
\begin{eqnarray}
\bm{\widetilde G}_{m>n} =
\begin{pmatrix}
0 & -g^{>}_{n, m} &-g^{<}_{n, m} &0\\
-g^>_{m, n} &0 &0 &g^<_{m, n}\\
-g^>_{m, n} &0 &0 &g^<_{m, n}\\
0 & g^{>}_{n, m} & g^{<}_{n, m} &0 
\end{pmatrix}.\nonumber 
\end{eqnarray} 
Using the relation
$$
g^{>,<}\left(t,~t'\right)= - g^{>,<,*}\left(t',~t\right) 
$$
yields Eqs.~\eqref{Delta_defn} and \eqref{Green's_function_matrix_defn1}.

\section{Derivation of the Temporal state overlap}
\label{app:overlap_derivation}
Deriving the path integral expression, given by Eq.~(\ref{eq:standard_path_integral}), includes inserting Grassmann resolutions of identity $\mathds{1}_\tau=\otimes_\sigma \mathds{1}_{\sigma,\tau},$ between each time-local operator---which can be either evolution operators $U_{\text{imp},\delta t}$, or combinations of operators such as $O_1 U_{\text{imp},\delta t}$---into Eq.~(\ref{eq:trace_expression}). 

For the combined impurity-bath system, the Grassmann identity reads
\begin{multline}
    \mathds{1}_{\sigma,\tau} =  \int d(\bar{\eta}_{\sigma,\tau},\eta_{\sigma,\tau})d(\bm{\bar{\xi}}_{\sigma,\tau},\bm{\xi}_{\sigma,\tau})\\ e^{-\bar{\eta}_{\sigma,\tau} \eta_{\sigma,\tau}- \bar{\bm{\xi}}_{\sigma,\tau}\bm{\xi}_{\sigma,\tau}}| \eta_{\sigma,\tau}, \bm{\xi}_{\sigma,\tau}\rangle \langle \bar{\eta}_{\sigma,\tau},\bar{\bm{\xi}}_{\sigma,\tau}|.
    \label{eq:GM_identity}
\end{multline} As introduced in the main text, $\bar{\eta}_{\sigma,\tau},\eta_{\sigma,\tau}$ are impurity variables, while $\bar{\bm{\xi}}_{\sigma,\tau}=(\bar{\xi}_{j=1,\sigma,\tau},\hdots,\bar{\xi}_{j=L,\sigma,\tau})^T,\,\bm{\xi}_{\sigma,\tau} = (\xi_{j=1,\sigma,\tau},\hdots,\xi_{j=L,\sigma,\tau})^T$ are the degrees of freedom of the environment  (consisting of $L$ fermionic modes). 
   
  In order to arrive at Eq.~(\ref{eq:expec_value_overlap}), we manipulate the path integral in a way that results in the following structure: All variables associated with the kernel of the spin-up (down) IF should be conjugate (non-conjugate) and opposite for the impurity kernel.
  This is achieved by making appropriate variable substitutions in the system-variables of the identity resolution, given by Eq.~(\ref{eq:GM_identity}). We define these modified identity resolutions as
  \begin{align}\label{eq:var_sub1}
      \mathds{1}^\prime_{\sigma,\tau} \quad &\text{ with substitution } \bar{\eta}_{\sigma,\tau}\to \eta_{\sigma,\tau},\, \eta_{\sigma,\tau} \to -\bar{\eta}_{\sigma,\tau},\\
      \label{eq:var_sub2}
   \mathds{1}^{\prime\prime}_{\sigma,\tau}\quad &\text{ with substitution } \bar{\eta}_{\sigma,\tau}\to -\eta_{\sigma,\tau},\, \eta_{\sigma,\tau} \to \bar{\eta}_{\sigma,\tau}.
  \end{align}
  With this, Grassmann identities are inserted between the hybridization and impurity evolution operators on the forward branch in the following way: 
  \begin{equation}
  U_{\text{imp},\delta t}\cdot\mathds{1}_{(2m+1)^+} \cdot U_{\text{hyb},\delta t}\cdot\mathds{1}_{(2m)^+},\end{equation}
  with the evolution operators as defined in Eq.~(\ref{eq:Trotter}). In this expression, we introduced
  \begin{align}
     \mathds{1}_{(2m+1)^+} =& \mathds{1}_{\uparrow,(2m+1)^+}\otimes \mathds{1}^\prime_{\downarrow,(2m+1)^+},\\
    \mathds{1}_{(2m)^+}=&  \mathds{1}^{\prime\prime}_{\uparrow,(2m)^+}\otimes \mathds{1}_{\downarrow,(2m)^+}.
  \end{align} On the backward branch, we insert identities as follows:
    \begin{equation}
 \mathds{1}_{(2m)^-} \cdot U^\dagger_{\text{hyb},\delta t}\cdot \mathds{1}_{(2m+1)^-} U^\dagger_{\text{imp},\delta t},\end{equation} with
 \begin{align}
     \mathds{1}_{(2m)^-} =& \mathds{1}_{\uparrow,(2m)^-}\otimes \mathds{1}^\prime_{\downarrow,(2m)^-},\\
     \mathds{1}_{(2m+1)^-} =& \mathds{1}^{\prime\prime}_{\uparrow,(2m+1)^-}\otimes \mathds{1}_{\downarrow,(2m+1)^-}.
  \end{align} 
  With these insertions, one arrives at Eq.~(\ref{eq:expec_value_overlap}) by following the standard text-book procedure for deriving the path integral. Note that these variable substitutions alter the signs of some components of the impurity kernel, while they amount to a simple renaming of variables for the IF. 
  
  In particular, the Grassmann kernel $\mathcal{D}^{O_1O_2}$ in Eq.~(\ref{eq:expec_value_overlap}) can formally be factorized as follows:

\begin{align}
    \nonumber
   &\mathcal{D}^{O_1O_2}[\{\bar{\eta}_{\downarrow,\tau},\eta_{\uparrow,\tau}\}]=\\\nonumber
   &  \mathcal{D}_{M^*}[\bar{\eta}_{\downarrow,(2M-1)^+},\eta_{\uparrow,(2M-1)^-}] \\\nonumber
   & \cdot\mathcal{D}_{(M-1)^+}[\bar{\eta}_{\downarrow,(2M-3)^+},\bar{\eta}_{\downarrow,(2M-2)^+},\eta_{\uparrow,(2M-3)^+},\eta_{\uparrow,(2M-2)^+}]\ldots & \\\nonumber
   & \cdot  \mathcal{D}_{n^+}[\bar{\eta}_{\downarrow,(2n-1)^+},\bar{\eta}_{\downarrow,(2n)^+},\eta_{\uparrow,(2n-1)^+},\eta_{\uparrow,(2n)^+}]\dots & &\hspace{-1cm}\\\nonumber
     & \cdot   \mathcal{D}_{0^*}[\bar{\eta}_{\downarrow,0^+},\bar{\eta}_{\downarrow,0^-},\eta_{\uparrow,0^+},\eta_{\uparrow,0^-}] \dots   & \hspace{-1cm}\\\nonumber
    & \cdot\mathcal{D}_{n^-}[\bar{\eta}_{\downarrow,(2n-1)^-},\bar{\eta}_{\downarrow,(2n)^-},\eta_{\uparrow,(2n-1)^-},\eta_{\uparrow,(2n)^-}]\dots &
    &\hspace{-1cm}\\
     & \cdot\mathcal{D}_{(M-1)^-}[\bar{\eta}_{\downarrow,(2M-3)^-},\bar{\eta}_{\downarrow,(2M-2)^-},\eta_{\uparrow,(2M-3)^-},\eta_{\uparrow,(2M-2)^-}].
      \label{eq:kernel_factor}
\end{align}
Here, the individual factors are the Grassmann kernels of the many-body operators from Eq.~(\ref{eq:Trotter}) and include the signs from the variable substitutions, given by Eqs.~(\ref{eq:var_sub1} and \ref{eq:var_sub2}), e.g., 
\begin{equation}\nonumber
     \mathcal{D}_{n^+} = \big\langle\bar{\eta}_{\downarrow,(2n)^+},-\eta_{\uparrow,(2n)^+}|U_{\text{imp},\delta t}|-\bar{\eta}_{\downarrow,(2n-1)^+},\eta_{\uparrow,(2n-1)^+}\big\rangle.
\end{equation}

\section{Derivation of the Grassmann kernels of the impurity gates}
\label{app:impurity_gate}

In order to evaluate any observable one must determine the Grassmann kernel of the corresponding many-body operator  in accordance with the order of the Grassmann variables in any single term (monomial) arising from the expansion of the influence functional given by Eq.~\eqref{eq_defn_IF}. In Eq. \eqref{eq:kernel_factor_main}, the kernel of the many-body operator is a product of kernels of the local impurity evolution operators.  
These kernels are sandwiched between  up-spin and down-spin IF-MPS and define a local map between the spin-up and spin-down Grassmann variables at a given time step. We follow a two-step procedure: (i) determine the signs in the individual gate-kernels, and (ii) adjust the non-local signs between the successive fermion operators. We consider the generic $4\times 4$ many-body gate to illustrate the procedure to calculate a Grassmann-kernel of the impurity gate.
\begin{enumerate}
\item {\it Forward branch impurity evolution operator.} The kernel of the many-body impurity evolution operator ($\tau>\tau'$) can be written generically as follows:   
\begin{eqnarray}
U =\vec{\bar{\eta}}_{\tau}^\mathrm{T}\cdot
\begin{bmatrix}
a_{0,0}&a_{0,1}&a_{0,2}&a_{0,3}\\
a_{1,0}&a_{1,1}&a_{1,2}&a_{1,3}\\
a_{2,0}&a_{2,1}&a_{2,2}&a_{2,3}\\
a_{3,0}&a_{3,1}&a_{3,2}&a_{3,3}\\
\end{bmatrix}\cdot
\vec{\eta}_{\tau '},
\label{Fwd_evol_gate}
\end{eqnarray}
where 
\begin{eqnarray}
\vec{\bar{\eta}}_\tau &=& \left(1, ~\bar{\eta}_{\uparrow, \tau}, ~\bar{\eta}_{\downarrow, \tau}, ~\bar{\eta}_{\downarrow, \tau}\bar{\eta}_{\uparrow, \tau}\right)^T,\\
\vec{\eta}_{\tau'} &=& \left(1, ~\eta_{\uparrow, \tau'}, ~\eta_{\downarrow, \tau'}, ~\eta_{\uparrow, \tau'}\eta_{\downarrow, \tau'}\right)^T.
\end{eqnarray}
We suppress the branch index in the Grassmann variables so as not to clutter the expressions and it is clear that the kernel only applies to the forward branch. 
We determine the signs in four steps:
\begin{enumerate}
\item Expand the kernel $U(\bar{\eta}_{\downarrow, \tau}, \bar{\eta}_{\uparrow, \tau}, \eta_{\uparrow, \tau'}, \eta_{\downarrow, \tau'})$  into the sum of Grassmann variable monomials.

\item Substitute variables to change the Grassmann integral into an overlap (same as Eqs.~\eqref{eq:var_sub1}, \eqref{eq:var_sub2}).

\item Re-order the Grassmann monomials according to the order of appearance of the Grassmann variables in the monomials of the IF-MPS overlap: (i) The barred Grassmann variables (corresponding to the down-spin IF-MPS) are to the left and the unbarred Grassmann variables (corresponding to the up-spin IF-MPS) are to the right of the gate-kernel, (ii) the Grassmann variables in the IF-MPS for a given spin-species appear in ascending order of the time index, so it implies that the Grassmann variables in the gate-kernel appear in descending order of the time index. 

As the impurity gate kernel will be appearing between the forward branch of a given time step (say $m$-th time step), the resulting order of the Grassmann variables in the impurity evolution kernel is 
\begin{eqnarray}
\label{eq:fwd_order}
U({\bar{\eta}_{\downarrow, 2m}},{\bar{\eta}_{\downarrow, 2m-1}},{\eta_{\uparrow, 2m}},{\eta_{\uparrow, 2m-1}}).
\end{eqnarray} 
Re-ordering the variables in each monomial results in changes in signs. Each swap between two Grassmann variables incurs a minus sign and if the total number of swaps needed to reach the order in Eq. \eqref{eq:fwd_order} is odd, the monomial is multiplied by minus sign.
\item  Re-express the monomial as a matrix, the local map between the up-spin and the down-spin variables. The forward gate-kernel for $U$ becomes
\begin{eqnarray}
\vec{\bar{\eta}}^\mathrm{T}_{\downarrow}\cdot
\begin{bmatrix}
a_{0,0}&-a_{1,0}&a_{0,1}&-a_{1,1}\\
a_{2,0}&-a_{3,0}&a_{2,1}&-a_{3,1}\\
-a_{0,2}&-a_{1,2}&a_{0,3}&a_{1,3}\\
-a_{2,2}&-a_{3,2}&a_{2,3}&a_{3,3}
\end{bmatrix}\cdot
 \vec{\eta}_{\uparrow},
\end{eqnarray}
where
\begin{eqnarray}
\vec{\bar{\eta}}_{\downarrow} &=& \left( 1, ~\bar{\eta}_{\downarrow, 2m},~ \bar{\eta}_{\downarrow, 2m-1}, ~\bar{\eta}_{\downarrow, 2m}\bar{\eta}_{\downarrow, 2m-1}\right)^\mathrm{T},\\
 \vec{\eta}_{\uparrow} &=& \left(1,~\eta_{\uparrow, 2m}, ~\eta_{\uparrow, 2m-1}, ~\eta_{\uparrow, 2m}\eta_{\uparrow, 2m-1}\right)^\mathrm{T}.
\end{eqnarray}
\end{enumerate}

\item {\it Backward branch impurity evolution operator.} 
The kernel of the many-body backward-evolution matrix is obtained in a similar way to the forward evolution gate: 
\begin{enumerate}
\item 
The many-body evolution operator $U^\dagger$ is obtained starting from Eq. \eqref{Fwd_evol_gate} as
\begin{eqnarray}
U^\dagger =\vec{\bar{\eta}}_{\tau '}^\mathrm{T}\cdot
\begin{bmatrix}
a^*_{0,0}&a^*_{1,0}&a^*_{2,0}&a^*_{3,0}\\
a^*_{0,1}&a^*_{1,1}&a^*_{2,1}&a^*_{3,1}\\
a^*_{0,2}&a^*_{1,2}&a^*_{2,2}&a^*_{3,2}\\
a^*_{0,3}&a^*_{1,3}&a^*_{2,3}&a^*_{3,3}\\
\end{bmatrix}\cdot
\vec{\eta}_{\tau},
 \label{Bkd_evol_gate}
\end{eqnarray}
where 
\begin{eqnarray}
\vec{\eta}_{\tau} &=& \left(1, ~\eta_{\uparrow, \tau}, ~\eta_{\downarrow, \tau}, ~\eta_{\uparrow, \tau}\eta_{\downarrow, \tau}\right)^\mathrm{T},\\
\vec{\bar{\eta}}_{\tau '} &=& \left(1, ~\bar{\eta}_{\uparrow, \tau '}, ~\bar{\eta}_{\downarrow, \tau '}, ~\bar{\eta}_{\downarrow, \tau '}\bar{\eta}_{\uparrow, \tau '}\right)^\mathrm{T}.
\end{eqnarray}
We suppress the branch index in the Grassmann variables so as not to clutter the expressions and it is clear that the kernel only applies to the backward branch.
\item We note that the Grassmann variables with smaller time-indices appear before variables with larger time-indices upon expanding the kernel. So, in order to return to the familiar set-up of the forward time-evolution case discussed above, shift the variables such that the unbarred variables (with larger time-index) appear before the barred variables (with smaller time-index). This results in some sign changes which is reflected in the kernel when re-expressed as a matrix:
\begin{eqnarray}
U^\dagger = 
\vec{\eta}_{\tau}^\mathrm{T}\cdot
 \begin{bmatrix}
 a^*_{0,0} & a^*_{0,1} &a^*_{0,2} &a^*_{0,3}\\
 a^*_{1,0} & -a^*_{1,1} &-a^*_{1,2} &a^*_{1,3}\\
 a^*_{2,0} & -a^*_{2,1} &-a^*_{2,2} &a^*_{2,3}\\
 a^*_{3,0} & a^*_{3,1} & a^*_{3,2} &a^*_{3,3}
 \end{bmatrix}\cdot
\vec{\bar{\eta}}_{\tau '}.
\end{eqnarray}
\item Proceeding from the above gate-kernel, we expand it to obtain the monomial expression and perform the variable substitution as in Eqs.~\eqref{eq:var_sub1}, \eqref{eq:var_sub2}. Finally the down-spin variables and up-spin variables are arranged into the described order of appearance as in Eq.~\eqref{eq:fwd_order}.  The signs incurred in the Grassmann monomials with the substitution and the rearrangement are absorbed into the matrix. Thus, one obtains the following backward gate-kernel for the operator $U^\dagger$: 
\begin{eqnarray}
\vec{\bar{\eta}}_{\downarrow}^\mathrm{T}\cdot
\begin{bmatrix}
a^*_{0,0} & a^*_{1,0} & -a^*_{0,1} & a^*_{1,1}\\
-a^*_{2,0} & a^*_{3,0} & -a^*_{2,1} & -a^*_{3,1}\\
a^*_{0, 2} & a^*_{1,2} & -a^*_{0,3} & a^*_{1,3}\\
a^*_{2,2} & -a^*_{3,2} & a^*_{2,3} & a^*_{3,3}
\end{bmatrix}\cdot
\vec{\eta}_{\uparrow},
\end{eqnarray}
where
\begin{eqnarray}
\vec{\bar{\eta}}_{\downarrow} &=& \left( 1, ~\bar{\eta}_{\downarrow, 2m},~ \bar{\eta}_{\downarrow, 2m-1}, ~\bar{\eta}_{\downarrow, 2m}\bar{\eta}_{\downarrow, 2m-1}\right)^\mathrm{T},\\
 \vec{\eta}_{\uparrow} &=& \left(1,~\eta_{\uparrow, 2m}, ~\eta_{\uparrow, 2m-1}, ~\eta_{\uparrow, 2m}\eta_{\uparrow, 2m-1}\right)^\mathrm{T}.
\end{eqnarray}

\end{enumerate}

\item {\it Global signs.}
The unbarred Grassmann variables in the monomial terms of the kernel expansion are in order with the Grassmann monomials of the up-spin IF-MPS. However, the barred Grassmann variables appear in the reverse order to the desired order of the down-spin IF-MPS. 
Note that the BCS structure of the IF imposes that the string of barred (spin down) variables and the string of unbarred (spin up) variables each have even parity.  Thus, reversing the string of $N$ barred variables introduces a sign ${(-1)}^{N/2} = i^{N}$. We take this into account in the
local kernels by including a factor of imaginary $i$ for each barred Grassmann variable in the kernels and it effectively counts the number of Grassmann variables. This procedure automatically produces a minus sign if $N/2 (mod ~2) = 1$ without the need to analyze the global string.

\item {\it Interleaving (Kronecker product of the evolution operators).} The
 Grassmann kernels of the forward and backward time-evolution operators appear
 as in the IF-MPS individually, but our organization of the variables in the
 IF-MPS is such that the Grassmann variables from the forward and backward
 branches are interleaved, i.e. $(\dots ~\eta_{2m, +}\eta_{2m, -}\eta_{2m+1,
 +}\eta_{2m+1, -}~\dots)$. Therefore, after doing the Kronecker product of the
 forward and backward kernels,  one needs to interleave the Grassmann kernels
 and keep track of the change in the signs due to exchanges of the Grassmann
 variables within the monomials generated from the Kronecker product. This
 results in the full time-evolution Grassmann kernel, which can be
 applied locally.
\end{enumerate}

\bibliography{manuscript_v2.bbl}

\end{document}